\documentclass{article}
\pdfoutput=1
\usepackage{xcolor,graphicx}
\usepackage[margin=1in]{geometry}
\usepackage{amsmath,amssymb,amsthm,mathtools,algorithm,algorithmic, comment}
\usepackage{newpxtext,newpxmath}
\usepackage{colortbl,rotating,booktabs,array}
\usepackage[hidelinks]{hyperref}
\usepackage[normalem]{ulem}

\mathtoolsset{showonlyrefs}

\newcommand\ZEROSENTINEL{\texttt{ZERO\_SENTINEL}}
\newcommand\ZEROMAXSENTINEL{\texttt{ZERO\_MAX\_SENTINEL}}
\newcommand\Fbf{F_{\texttt{bf16}}}

\title{Providing Accuracy Guarantees for the Fixed-Point Emulation of Matrix Multiplication}
\author{Cole Brower\\{\normalsize NVIDIA Corporation}\\{\normalsize Santa Clara, California, USA}\\{\normalsize cbrower@nvidia.com} \and Peter Caday\\{\normalsize NVIDIA Corporation}\\{\normalsize Santa Clara, California, USA}\\{\normalsize pcaday@nvidia.com} \and John Gunnels\\{\normalsize NVIDIA Corporation}\\{\normalsize Santa Clara, California, USA}\\{\normalsize jgunnels@nvidia.com}}
\date{November 13, 2026}

\newcommand{\ZZ}{\mathbb Z}
\newcommand{\RR}{\mathbb R}

\newcommand\V\mathbf

\newcommand{\emax}{e^{(\max)}}
\newcommand{\emin}{e^{(\min)}}

\let\eps\epsilon

\DeclareMathOperator\fma{fma}

\DeclareMathOperator\Var{Var}

\DeclareMathOperator\esc{esc}

\newcommand\set[2]{\{#1:#2\}}
\DeclarePairedDelimiter\floor\lfloor\rfloor
\DeclarePairedDelimiter\ceil\lceil\rceil

\DeclarePairedDelimiter\abs\lvert\rvert
\DeclarePairedDelimiter\norm\lVert\rVert

\newcommand\sinfty{{\infty\vphantom1}}

\newtheorem{theorem}{Theorem}[section]

\newtheorem{proposition}[theorem]{Proposition}

\newcommand{\Einf}{E_{\infty}}

\begin{document}

\maketitle

\begin{abstract}
Fixed-point methods such as Ozaki-II emulate double-precision matrix multiplication (GEMM) on fast low-precision hardware. Software that chooses the mantissa bit count automatically is numerically safe when the emulated product keeps the error guarantee of the floating-point GEMM it replaces. We analyze the worst-case and average-case error of fixed-point and floating-point dot products, define exponent span capacities that determine the extra mantissa bits an input requires, and approximate them with a low-precision GEMM. We prove that 53 bits plus such a capacity and a small constant preserve the classical worst-case error bound for a floating-point dot product, and we demonstrate this guarantee in software on several test cases. In a developer build of cuBLAS on an NVIDIA B200, each implemented bit-count rule earns grade~A on Test~2c of ``Grading the BLAS'' at every size and exponent range tested, falling back to double precision at the widest ranges; in exact arithmetic, the bound holds on every adversarial and random input tested.
\end{abstract}

\section{Introduction}\label{sec:intro}

On modern GPUs, low-precision general matrix multiplication (GEMM) outperforms traditional IEEE-754 single- and double-precision arithmetic by many orders of magnitude, driven by the computational demands of artificial intelligence. Recently-introduced algorithms, in particular the Ozaki-I and Ozaki-II schemes~\cite{ozaki2012error, ootomo2024dgemm, ozaki2026scheme2}, leverage this trend, using fast low-precision hardware to accelerate higher-precision GEMMs. In both algorithms, floating-point input matrices are scaled and quantized to integer matrices, and the resulting high-precision integer GEMM is reduced to a series of lower-precision integer GEMMs. A final reconstruction step undoes the scaling to produce a final floating-point product. We denote this class of algorithms \emph{fixed-point methods}.

Fixed-point methods' accuracy is primarily driven by the precision of the integer quantization (fixed-point mantissa bit count), which is a tunable parameter. A natural question is how to choose the mantissa bit count for a particular application, and in particular, how to match the accuracy of floating-point GEMM. As the accuracy of floating-point GEMM is itself an involved topic~\cite{higham2002accuracy, HighamMary, demmel2026grade}, this is a subtle question. In a previous paper~\cite{schwarz2026guaranteed}, the \emph{exponent span capacity} (ESC) was proposed as a measure of the number of additional bits of integer precision beyond the 53 mantissa bits of native FP64 values. Fasi et al.~\cite{fasi2026sharper} recently proved that this choice (with a fixed number of additional bits) ensures that the \emph{worst-case} fixed-point error matches the \emph{worst-case} floating-point GEMM error.

In this paper, we carry the accuracy analysis further, analyzing both average-case and worst-case error for fixed-point and floating-point dot products. Our results naturally lead to two new variants of ESC that provide tighter bounds on the number of required fixed-point mantissa bits to match floating-point accuracy. We show that these variants can be efficiently approximated using a low-precision (bfloat16) GEMM, and also derive an efficient new approximation algorithm for the original ESC that is more robust and often provides tighter ESC bounds than the coarsened approximation of~\cite{schwarz2026guaranteed}. Finally, we discuss a new approach to scaling in Ozaki-II and demonstrate that our implementations pass known GEMM accuracy tests from Demmel et al~\cite{demmel2026grade}, as well as additional proposed stress tests which demonstrates the robustness of our ESC algorithms.

\section{Dot Product Accuracy}\label{sec:accuracy-theory}

In this section, we present worst-case and average-case error analyses for both fixed-point and floating-point dot products. Our assumption throughout is that integer and fixed-point arithmetic is performed exactly, so that the only errors in fixed-point methods arise from discretization. Note that this is true of the Ozaki-II scheme but not of the original Ozaki-I scheme, where some lower-order contributions to the integer product are not computed. At the cost of reduced performance, Ozaki-I may be readily modified to compute the integer product in full precision, in which case our analysis applies.

We will also consider the fundamental uncertainty of the dot product, that is, error due to pre-existing rounding of the floating-point inputs, which is independent of the dot product algorithm and reflects the numeric conditioning of the dot product itself.

For clarity, we collect proofs for all the results of this section in Section~\ref{s:proofs}.

\subsection{Definitions}
\label{s:accuracy-definitions}

\paragraph{Floating-Point Numbers}
Fix a floating point domain $F$ with $M$ mantissa bits, including the leading 1. For a nonzero $a\in F$, let $\eps(a)$ be the maximum distance a number can be from $a$ and still round to $a$. We will assume round-to-nearest throughout, so $\eps(a) = 2^{\floor{\log_2\abs{a}} - M}$ for normal numbers, with no assumptions on the tie-breaking rule. For all $a$, we have $\frac12 \abs a\cdot\eps(1)<\eps(a)\leq \abs a\cdot\eps(1)$. Finite normal floating-point numbers are arranged in \emph{binades} $[2^e, 2^{e+1})$, where $e$ is an integer exponent.

\paragraph{Fixed-Point Numbers}
A fixed-point number with $\hat M$ mantissa bits relative to a maximum value $A$ is a number of the form $\pm 2^{\floor{\log_2\abs{A}} - \hat M + 1}\cdot m$, where $m$ is an integer in $[-2^{\hat M}, 2^{\hat M}]$. For any $a\in[-\abs A,\abs A]$, the distance between $a$ and the nearest fixed-point number is at most $\hat\eps(A) = 2^{\floor{\log_2 \abs A} - \hat M}$. (For $A\in F$ and $\hat M\geq M$, the values $m=\pm 2^{\hat M}$ are never needed.) We will often think of $\hat M$ as a sum $M+E$, where $E$ is a nonnegative number of additional fixed-point mantissa bits.

\medskip
\noindent We will use the following variables throughout:
\begin{itemize}
\item $x,\,y\in F^k$: input floating-point vectors, of length $k$.
\item $z=x\cdot y\in\RR$: mathematically exact dot product.
\item $w = x\circ y\in\RR^k$: Hadamard product, $w=(w_i) = (x_iy_i)$.
\item $\hat x,\,\hat y\in \RR^k$: fixed-point approximations to $x$, $y$, relative to $\norm x_\infty$ and $\norm y_\infty$, respectively.
\item $\hat z=\hat x\cdot \hat{y}$: mathematically exact dot product of $\hat x$ and $\hat y$.
\end{itemize}

For simplicity, we will assume throughout that the inputs $x$, $y$ and intermediate floating-point values are finite, normal, nonzero numbers. Denormal numbers, underflow, and overflow will reduce accuracy of floating-point algorithms, but not fixed-point algorithms, so this simplification is reasonable. Without loss of generality, we can drop zero inputs without affecting any algorithm's accuracy.

For average-case analysis, we need to choose probability distributions from which to draw inputs. Typically we will characterize rounding and discretization errors as belonging to uniform distributions, and let $\mathcal U(a,b)$ denote the uniform distribution over the interval $[a,b]$.

\subsection{Input Rounding Error Bounds}
\label{s:uncertainty-stmt}

In general, the floating-point inputs $x$, $y$ originate from a previous floating-point operation, where they were rounded from some exact real values $\overline x,\,\overline y\in\RR^k$. Because we do not know the true values $\overline x$, $\overline y$, while we can compute the exact product $z\in\RR$, there is a fundamental uncertainty that limits the useful precision of $z$.

While it is well-known how input rounding affects the exact product $z$, it will be useful to compare these bounds (which represent an ``error floor'' for other algorithms) against our later fixed- and floating-point error bounds.

\begin{proposition}
\label{p:input-rounding-worst}
The error due to input rounding is
\[
\abs{\overline x\cdot\overline y - z} \leq  2\eps(1)\cdot\norm w_1 + O(\eps(1)^2).
\]
\end{proposition}

To quantify average-case error, replace $\overline x$, $\overline y$ by vector-valued random variables $X=(X_i)$, $Y=(Y_i)$ with entries drawn from independent uniform distributions:
\[
X_i \sim \mathcal U(x_i-\eps(x_i), x_i+\eps(x_i)),
\qquad
Y_i \sim \mathcal U(y_i-\eps(y_i), y_i+\eps(y_i)).
\]
We can then examine the standard deviation of $X\cdot Y$:

\begin{proposition}
\label{p:input-rounding-average}
The standard deviation due to input rounding is on the order of $\eps(1)\cdot\norm w_2$. More specifically,
\[
\frac{1}{\sqrt{6}}\cdot \eps(1)\cdot\norm w_2
\leq
\sigma(X\cdot Y-z)
\leq
\frac{2}{\sqrt{6}}\cdot \eps(1)\cdot\norm w_2 + O(\eps(1)^2).
\]
\end{proposition}

Straightforward proofs of both results may be found in Section \ref{s:uncertainty-proof}.

\subsection{Fixed-Point Error Bounds}
\label{s:fixed-point-error-stmt}

The analysis for fixed-point error is similar to input rounding, except that we replace input rounding errors $\eps(x_i)$, $\eps(y_i)$ with discretization errors $\hat\eps(\norm x_\sinfty)$, $\hat\eps(\norm y_\sinfty)$, which depend only on the magnitude of the largest elements in $x$ and $y$. In addition, when $\hat M\geq M$, the largest elements in $x$, $y$ can be exactly represented in fixed point, and do not contribute any error to $\hat z$, allowing tighter bounds.

\begin{proposition}
\label{p:fixed-worst}
The error of the fixed-point dot product $\hat z$ is bounded by
\begin{align}
\abs{\hat z - z}
	&\leq \hat\eps(1)\cdot\left[\norm x_\sinfty\norm y_1 + \norm y_\sinfty\norm x_1\right] + O(\hat\eps(1)^2)\\
	&\leq 2\hat\eps(1)\cdot k\norm x_\sinfty\norm y_\sinfty + O(\hat\eps(1)^2).
\end{align}
Furthermore, if $E=\hat M-M>0$, the bound can be made twice as tight:
\[
\abs{\hat z - z}\leq \hat\eps(1)\cdot k\norm x_\sinfty \norm y_\sinfty + O(\hat\eps(1)^2).
\]
\end{proposition}

For average-case error, we model the discretization errors as independent uniform random variables:
\begin{align}
\label{e:fxp-model}
\hat x_i - x_i &\sim \mathcal U(-\hat\eps(\norm x_\sinfty), \hat\eps(\norm x_\sinfty)),
&
\hat y_i - y_i &\sim \mathcal U(-\hat\eps(\norm y_\sinfty), \hat\eps(\norm y_\sinfty)).
\end{align}
Note that this probabilistic model, while a relatively mild assumption on the input vectors, may not be sufficiently representative of all applications.

\begin{proposition}
\label{p:fixed-avg}
Under the independent rounding model~\eqref{e:fxp-model}, $\mathbb E[\hat z]=z$, and the standard deviation of $\hat z$ is on the order of $\hat\eps(1)\sqrt{\norm x_2^2\norm y_\sinfty^2 + \norm y_2^2\norm x_\sinfty^2}$. More precisely,
\begin{align}
\label{e:fxp-avg}
\frac{\hat\eps(1)}{2\sqrt 3} \sqrt{\norm x_2^2\norm y_\sinfty^2 + \norm y_2^2\norm x_\sinfty^2}
\leq
\sigma(\hat z)
\leq
\frac{\hat\eps(1)}{\sqrt 3} \sqrt{\norm x_2^2\norm y_\sinfty^2 + \norm y_2^2\norm x_\sinfty^2} + O(\hat\eps(1)^2).
\end{align}
Furthermore,
\begin{align}
\label{e:fxp-avg-inf}
\sigma(\hat z)
\leq
\sqrt{\frac23}\hat\eps(1)\cdot \sqrt k \norm x_\sinfty\norm y_\sinfty + O(\hat\eps(1)^2).
\end{align}
\end{proposition}
Proofs may be found in Section~\ref{s:fixed-point-error-proof}.

\subsection{Floating-Point Error Bounds}
\label{s:fp-error-stmt}

The study of floating-point dot product accuracy has a long history in the literature. A straightforward application of the IEEE-754 rounding rules leads to a $O(k/(1-k\eps(1)))$ worst-case bound~\cite{higham2002accuracy}, which was sharpened to $O(k)$ by Jeannerod and Rump~\cite{JeannerodRump}. However, it is well-recognized that the worst-case bounds significantly overestimate the average error. Higham and Mary~\cite{HighamMary} prove that average error, under a randomized rounding error model, grows as $O(\sqrt{k\log k})$, much slower than the worst-case $O(k)$ bound. In this section, we pursue a different approach to average error, calculating variances instead of using a concentration inequality. Furthermore, we will allow accumulation order to vary, rather than analyzing a specific order of floating-point operations.

We start with a basic sequential dot product implementation, assuming hardware with fused multiply-add (FMA):
\[
z \approx d(x,y) = \fma(\cdots\fma(\fma(0, x_1, y_1), x_2, y_2) \cdots, x_k, y_k).
\]
Here $\fma(a, b, c)$ computes $a+bc$ correctly rounded to an element of $F$.
It is not hard to see that the rounding error experienced by the $i^{\text{th}}$ FMA in this implementation is bounded (to first order) by $\eps(s_i)$, where $s_i=\sum_{j=1}^i x_jy_j=\sum_{j=1}^i w_j$ is the $i^{\text{th}}$ partial sum. More precisely, if $e_i$ is the rounding error incurred by the $i^{\text{th}}$ FMA, then
\begin{equation}
d(x,y) = x_1y_1 + e_1 + x_2y_2 + e_2 + \dotsc + x_ky_k + e_k = z + \sum_{i=1}^k e_i,
\end{equation}
with errors bounded by
\begin{equation}
\abs{e_i} \leq \eps\left(\sum_{j=1}^{i}x_jy_j + \sum_{j=1}^{i-1}e_j\right)
= \eps\left(s_i + \sum_{j=1}^{i-1}e_j\right).
\end{equation}
In almost all cases, the accumulated error $\sum_{j=1}^{i-1}e_j$ does not push $s_i$ into a higher binade, in which case we may simplify the right-hand side to $\abs{e_i}\leq \eps(s_i)$.

We first recall Jeannerod and Rump's worst-case error bound, which in our setting reads
\begin{equation}
\label{e:fp-worst-jr}
\abs{d(x,y)-z} \leq \eps(1)\cdot k\norm w_1.
\end{equation}
Note that this bound does not assume the use of FMA.

Now consider the variance under the model that each rounding error is drawn independently: $e_i\sim\mathcal U(-\eps(s_i),\eps(s_i))$. As Higham and Mary~\cite{HighamMary} discuss at length, it is \emph{not} necessarily safe to assume that rounding errors are independent, but nevertheless such a model is expected to provide a reasonably accurate proxy for the correct distribution.

Under this model, we have
\begin{align}
\label{e:pfpdp-variance-pre}
\Var(d(x,y)) &= \sum_{i=1}^k\Var(e_i) = \frac13\sum_{i=1}^k \eps(s_i)^2 \approx \frac{\eps(1)^2}{3} \norm s_2^2.
\end{align}
Using the fact that $\frac12<\eps(s_i)/\abs{s_i}\eps(1)\leq 1$, we conclude that
\begin{align}
\label{e:pfpdp-variance}
\Var(d(x,y)) &= \frac{c\eps(1)^2}{3} \norm s_2^2,
&&
\text{for some } c\in\left(\tfrac14, 1\right].
\end{align}
Not surprisingly, the 2-norm of the partial sums $s_i$ is the determining factor: the larger the accumulator values over the course of the computation, the more error can be expected.

So far, we have accumulated the terms $x_iy_i$ in the order $1,\dotsc,k$, but dot product implementations in practice do not always follows this ordering, and it seems reasonable to allow a dot product to accumulate in any order desired. Given a permutation $\pi\in S_k$, define the permuted-sequential FMA dot product
\begin{equation}
d_\pi(x,y) = \fma(\cdots\fma(\fma(0, x_{\pi(1)}, y_{\pi(1)}), x_{\pi(2)}, y_{\pi(2)}) \cdots, x_{\pi(k)}, y_{\pi(k)}).
\end{equation}
As the permutation varies, so does the partial-sum norm $\norm s_2$: we can make $\norm s_2$ larger by moving large entries of $w$ having the same sign to the beginning of the dot product, or smaller, by moving large entries of $w$ to the end and arranging for cancellation between neighboring terms when possible.

Using these ideas, we can derive a lower bound for the worst-case variance $\Var(d_\pi(x,y))$ over all possible permutations $\pi$, and also characterize the average variance over all permutations.

\begin{proposition}
\label{p:fp-worst}
For any $x$, $y$, under the independent rounding error model,
\begin{align}
\max_{\pi\in S_k} \Var(d_\pi(x,y)) &\geq \frac{c\eps(1)^2}{96} k\norm w_1^2
\end{align}
for some $c\in(\tfrac14, 1]$.
\end{proposition}

\begin{proposition}
\label{p:fp-avg}
Under the independent rounding error model, the average variance over all permutations is given by
\begin{align}
\frac{1}{k!}\sum_\pi\Var(d_\pi(x,y)) = \frac{c\eps(1)^2}{18}\cdot(k+1)\left(2z^2+\norm w_2^2\right),
\end{align}
for some $c\in(\tfrac14,1]$.
\end{proposition}
Proofs may be found in Section~\ref{s:fp-error-proof}.

\subsection{Error Bound Summary}

Table~\ref{t:bounds} collects the bounds of this section in their simplified forms. Effectively, the error bounds split into two families: the fixed-point bounds depending on $\norm x_\sinfty \norm y_\sinfty$, and the other bounds depending on $\norm w_1$ and $\norm w_2$. Later, we will define a family of \emph{ESC} (exponent span capacity) functions that capture the gaps between these vector norms. Note that none of the bounds are affected by cancellation in the summation, with the exception of the $z^2$ term in the floating-point average order bound. In the absence of cancellation, $\abs z=\norm w_1$.

The other major difference between bounds are the factors of $k$ found in the fixed- and floating-point bounds, but absent from the input rounding bounds, which represent an effective accuracy limit for any dot product. This will imply that fixed-point methods allow us to achieve the best possible error bounds, at a cost of $\Theta(\log_2 k)$ extra mantissa bits.

\begin{table}[tb]
\centering\footnotesize\renewcommand{\arraystretch}{1.25}\setlength{\tabcolsep}{5pt}
\caption{Error bounds, in terms of the norms of $x$, $y$, and $w = x \circ y$, to first order in $\eps(1)$ and $\hat\eps(1)$. For brevity, $C_1, C_2, \dotsc$ represent numbers that may depend on $x$, $y$ but lie within a range bounded independently of $x$, $y$. Variances assume the uniform independent rounding models of this section.}\label{t:bounds}
\bigskip
\begin{tabular}{@{}l l l@{}}\toprule
Error source & Worst case & Variance
\\ \midrule
Input rounding & $C_1\eps(1)\cdot\norm w_1$ & $C_2\eps(1)^2\cdot\norm w_2^2$ \\
Fixed point & $C_3\hat\eps(1)\cdot k\norm x_\sinfty \norm y_\sinfty$& $C_4\hat\eps(1)^2\cdot k\norm x_\sinfty^2\norm y_\sinfty^2$ \\
Floating point, averaged over all orders & $\eps(1)\cdot k\norm w_1$ & $C_5\eps(1)^2\cdot(k+1)(2z^2 + \norm w_2^2)$ \\
Floating point, worst order & $\eps(1)\cdot k\norm w_1$ & $C_6\eps(1)^2\cdot k\norm w_1^2$ \\
\bottomrule\end{tabular}
\\\bigskip
\begin{tabular}{r cccccc}
\toprule
&$C_1$&$C_2$&$C_3$&$C_4$&$C_5$&$C_6$
\\\midrule
Minimum value & 2 & $\frac16$ & -                        & -         & $\frac1{72}$ & $\frac1{384}$\\
Maximum value & 2 & $\frac23$ & 1 ($E>0$), 2 ($E\leq 0$) & $\frac23$ & $\frac1{18}$ & $\frac1{96}$\\
\bottomrule
\end{tabular}
\end{table}

\section{Achieving Floating-Point-Like Accuracy}\label{s:accuracy-matching}

Equipped with the dot product accuracy bounds of Section~\ref{sec:accuracy-theory}, our goal is to choose a fixed-point mantissa count $\hat M$ given $M$, $x$, and $y$ so that fixed-point methods provide comparable accuracy to a standard floating-point implementation. Fasi, Mary, and Mikaitis~\cite{fasi2026sharper} recently performed a similar analysis, using worst-case bounds. We will extend their analysis to average-case bounds and sharpen their results for worst-case bounds. In the process, it will be natural to introduce a new family of $p$-exponent span capacity ($\esc_p$) functions, which given $x$, $y$ will determine the necessary number of extra mantissa bits $E=\hat M - M$. In particular, $\esc_\infty$ is identical to the original ESC function, as defined in~\cite{schwarz2026guaranteed}, up to a small constant number of bits. By taking a maximum over the dot products within a GEMM, these results immediately generalize to full matrix multiplications.

\subsection{Matching Fixed- and Floating-Point Bounds}
\label{s:matching-bounds}

\paragraph{Matching Worst-Case Error Bounds}

Referring to Table~\ref{t:bounds} and recalling that $\eps(1)=2^{-M}$, $\hat\eps(1)=2^{-\hat M}$, the fixed-point worst-case error is no larger than the floating-point worst-case error when
\[
C_3 2^{-\hat M}\norm x_\sinfty\norm y_\sinfty \leq 2^{-M}\norm w_1,
\]
or equivalently,
\begin{equation}
\label{e:worst-match-pre}
E \geq \log_2\frac{\norm x_\sinfty\norm y_\sinfty}{\norm w_1} + \log_2(C_3).
\end{equation}
Recalling that we may take $C_3 = 1$ if $E>0$ and $C_3=2$ otherwise, we may weaken this to
\begin{equation}
\label{e:worst-match}
E \geq E_{\text{worst}} = \max\left(1,\log_2\frac{\norm x_\sinfty\norm y_\sinfty}{\norm w_1}\right).
\end{equation}
Equation~\eqref{e:worst-match} can be thought of as a sharpened version of Fasi et al.'s ESC bound~\cite{fasi2026sharper}.

We may also wish to pick $E$ large enough to achieve a stronger bound, where the fixed-point error is no worse than the error due to input rounding, in the worst case. This scenario requires
\[
C_3 2^{-\hat M}k\norm x_\sinfty\norm y_\sinfty \leq C_1 2^{-M}\norm w_1.
\]
By similar reasoning as above, and recalling $C_1=2$, it suffices to guarantee
\begin{equation}
\label{e:worst-match-ideal}
E \geq \overline E_{\text{worst}} = \max\left({1},\log_2\frac{\norm x_\sinfty\norm y_\sinfty}{\norm w_1} + \log_2 k - 1\right).
\end{equation}
In other words, we may pay $\log_2 k$ extra mantissa bits in order to achieve optimal fixed-point accuracy (\emph{optimal} in the sense of matching the inevitable worst-case error due to input rounding, which captures the ill-conditioning of the dot product).

\paragraph{Matching Average-Case Error Bounds}

If we average over all floating-point accumulation orders, fixed-point variance will be no larger than floating-point variance when
\begin{equation}
\label{e:avg-match-raw}
C_4 2^{-2\hat M} \norm x_\sinfty^2\norm y_\sinfty^2 \leq C_5\left(1+\frac1k\right)2^{-2M}(2z^2+\norm w_2^2).
\end{equation}
In the presence of cancellation, the term $2z^2$ could be, in the worst case, zero, so we may focus on the 2-norm of $w$. Dropping the $z$ term, we are assured fixed-point variance is no greater than floating-point variance if
\begin{equation}
\label{e:avg-match-pre}
E \geq \log_2\frac{\norm x_\sinfty\norm y_\sinfty}{\norm w_2} + \frac12\log_2\frac{C_4k}{C_5(k+1)}.
\end{equation}
Weakening this slightly to account for the bounds on $C_4$, $C_5$, it suffices to ensure
\begin{equation}
\label{e:avg-match}
E \geq E_{\text{avg}} =\log_2\frac{\norm x_\sinfty\norm y_\sinfty}{\norm w_2} + 2 + \log_2(\sqrt{3}).
\end{equation}
On the other hand, if there is no cancellation, $\abs z=\norm w_1$, and it suffices to ensure the weaker inequality
\begin{equation}
\label{e:avg-match-no-cancel}
E \geq E_{\text{avg}}' =\log_2\frac{\norm x_\sinfty\norm y_\sinfty}{\norm w_1} + 2 + \log_2(\sqrt{3/2}).
\end{equation}
Finally, if we wish to match the noise floor given by input rounding, instead of matching floating-point accuracy, we have
\[
C_4 2^{-2\hat M} k\norm x_\sinfty^2\norm y_\sinfty^2 \leq C_2 2^{-2M}\norm w_2^2
\]
and given $C_4\leq \frac23$, $C_2\geq\frac16$, it suffices to ensure
\begin{equation}
\label{e:avg-match-ideal}
E \geq \overline E_{\text{avg}} = \log_2\frac{\norm x_\sinfty\norm y_\sinfty}{\norm w_2} + \frac12\log_2 k + 1.
\end{equation}

\subsection{The \protect{$p$}-Exponent Span Capacity}
\label{s:p-esc}

\noeqref{e:worst-match}
\noeqref{e:worst-match-ideal}
\noeqref{e:avg-match}
\noeqref{e:avg-match-no-cancel}
\noeqref{e:avg-match-ideal}

Examining the five options~(\ref{e:worst-match}--\ref{e:avg-match-ideal}) for choosing the fixed-point mantissa width, it is natural to introduce the \emph{$p$-exponent span capacity}
\begin{align}
\esc_p(x,y) &= \log_2\frac{\norm x_\sinfty\norm y_\sinfty}{\norm w_p}.
&&
(1\leq p\leq\infty)
\end{align}
Evidently $E_{\text{worst}} = \max(1, \esc_1(x,y))$ and $E_{\text{avg}}' = \esc_1(x,y)+O(1)$, while $E_{\text{avg}} = \esc_2(x,y)+O(1)$. The ideal bounds $\overline E_{\text{worst}}, \overline E_{\text{avg}}$ similarly can be represented in terms of $\esc_1$, $\esc_2$, plus an additional $\Theta(\log_2 k)$ overhead. Note that the original ESC function of Schwarz et al.~\cite{schwarz2026guaranteed} is given by
\begin{align}
\esc(x,y) &= \floor*{\log_2\norm x_\sinfty} + \floor*{\log_2\norm y_\sinfty} - \floor*{\log_2\norm w_\sinfty}\\
&\in [\esc_\sinfty(x,y) - 2, \esc_\sinfty(x,y)+1].
\end{align}
Accordingly, we can often use the original $\esc$ and the new $\esc_\infty$ interchangeably, with the proviso that they may differ by up to 2 bits.

Since $\norm w_p \geq \norm w_{p'}$ whenever $p\leq p'$, we immediately have
\[
\esc_1(x,y) \leq \esc_2(x,y)\leq \esc_\sinfty(x,y),
\]
and we conclude that the original ESC is overly conservative for matching either worst- or average-case error, where $\esc_1$ and $\esc_2$ respectively suffice. This has also been observed experimentally, cf.~\cite{caday2026twoM}.

The $p$-ESC extends readily to matrix multiplications. Given $A$, $B$, let $a_i$ denote the $i^{\text{th}}$ row of $A$ and $b_j$ the $j^{\text{th}}$ column of $B$.
Since all fixed-point accuracy bounds are monotonic with respect to $\hat M$, it suffices to ensure that the worst dot product $a_i\cdot b_j$ is calculated in sufficient accuracy. Accordingly, define
\[
\esc_p(A,B)=\max_{i,j}\esc_p(a_i,b_j).
\]

\section{Computing $p$-Exponent Span Capacity}\label{sec:algorithms}

Having defined $p$-exponent span capacity and shown how it can be used to derive fixed point mantissa bit widths, we now present two new algorithms for efficiently approximating $\esc_p$. Both algorithms are built on low-precision (bfloat16) matrix multiplication, which is very fast on modern hardware.

\subsection{Computing $\infty$-Exponent Span Capacity}\label{sec:alg-esc}\label{sec:alg-binned}

We begin by recalling Schwarz et al.'s \emph{coarsened ESC} algorithm~\cite{schwarz2026guaranteed}, which provides an upper bound for $\esc_\infty(x,y)$, and discussing its weaknesses. We then present a new \emph{binned ESC} algorithm that avoids these weaknesses while remaining efficient to compute on modern hardware.

The algorithms we discuss here will in fact calculate upper bounds for Schwarz et al's original ESC function, $\esc(x,y)$. As mentioned in Section~\ref{s:p-esc}, $\esc(x,y)$ and $\esc_\infty(x,y)$ differ by at most 2 bits. We will let $e(a) = \floor{\log_2 \abs a}$ denote the exponent of a nonzero floating-point number $a\in F$, and write $e_x = \max_i e(x_i)$ and $e_y=\max_i e(y_i)$.

\subsubsection{Coarsened ESC}

In \cite{schwarz2026guaranteed}, the authors
observe that it is computationally expensive to calculate $\esc(x,y)$ directly. To reduce runtime, they \emph{coarsen} the data in the $k$ dimension. Specifically, for a coarsening factor $c$ and $x, y \in F^k$, partition the indices into blocks
$B_j = \{(j-1)c+1, \dots, \min(jc, k)\}$ $(j = 1, \dots, \ceil{k/c})$, and keep only the extreme exponents of each block,
\[
\emax_j(x) = \max_{i \in B_j} e(x_i), \qquad
\emin_j(x) = \min_{i \in B_j} e(x_i),
\]
and likewise for $y$. Now define a coarsened ESC
\begin{equation}\label{eq:esc-coarsened}
  \Einf^{(c)}(x, y) = e_x + e_y - \max_j \max\bigl(
    \emax_j(x) + \emin_j(y),\;
    \emin_j(x) + \emax_j(y)\bigr)
\end{equation}

Within a block, the entry attaining $\emax_j(x)$ is paired with an entry of $y$ of exponent at least $\emin_j(y)$, so $\Einf^{(c)} \ge \Einf^{(1)} \ge \esc$, and since $\|x\|_\infty \|y\|_\infty < 2^{e_x + e_y + 2}$,
\[
\esc_\infty < \Einf^{(c)} + 2.
\]

Zero inputs represent a special case in the coarsened ESC algorithm, because the exponent is effectively $-\infty$. The challenge is that once we compress the inputs to a minimum and maximum exponent per block, we cannot know how the zero values of this block would interact with other blocks it multiplies against: the dot product of two sparse blocks may be zero (no contribution to ESC) or nonzero (in which case coarsened ESC provides no information).

Accordingly, Schwarz et al.~\cite{schwarz2026guaranteed} propose representing zeros with two sentinels: $e(0) = \ZEROSENTINEL$ and $e_x,\,e_y = \ZEROMAXSENTINEL$ if $x$, $y$ are all zero, respectively. The sentinel values are both far below the exponent range of $F$ and are chosen such that \eqref{eq:esc-coarsened} produces a zero value for $\Einf^{(c)}$ when either $x$ or $y$ contain all zeros, and otherwise a very large sentinel value when every block contains zeros, indicating a failure to bound $\Einf$. Their ADP framework flags these large sentinel ESC values and dispatches to native floating-point computation.

\begin{algorithm}[t]
\caption{\texttt{coarsened\_quantize}$(x)$}\label{alg:coarsened_quantize}
\begin{algorithmic}[1]
\REQUIRE $x \in F^k$ finite; coarsening factor $c$; $\ZEROSENTINEL = -16384$, $\ZEROMAXSENTINEL = -32768$
\ENSURE $\emax(x), \emin(x) \in \ZZ^{\ceil{k/c}}$, the largest and smallest exponent of each block; $e_x$, the largest exponent of $x$
\STATE $e(x_i) \gets \floor{\log_2 \abs{x_i}}$ if $x_i \neq 0$, else \ZEROSENTINEL
\FOR{$j = 1, \dots, \ceil{k/c}$}
\STATE $B_j \gets \{(j-1)c + 1, \dots, \min(jc,\, k)\}$
\STATE $\emax_j(x) \gets \max_{i \in B_j} e(x_i)$
\STATE $\emin_j(x) \gets \min_{i \in B_j} e(x_i)$
\ENDFOR
\STATE $e_x \gets \max_j \emax_j(x)$
\IF{$e_x = \ZEROSENTINEL$}
\STATE $e_x \gets \ZEROMAXSENTINEL$
\ENDIF
\RETURN $\emax(x), \emin(x), e_x$
\end{algorithmic}
\end{algorithm}

\begin{algorithm}[t]
\caption{Coarsened $\esc_\infty$ calculation for $x, y \in F^k$ with coarsening factor $c$}\label{alg:esc_coarsened}
\begin{algorithmic}[1]
\REQUIRE $x, y \in F^k$ finite; coarsen factor $c$; $\ZEROSENTINEL = -16384$, $\ZEROMAXSENTINEL = -32768$
\ENSURE the count $\hat M = 53 + \Einf^{(c)} + 1$ with $\Einf^{(c)} \ge \Einf(x, y)$, or the fallback count
\STATE $\emax(x), \emin(x), e_x \gets \texttt{coarsened\_quantize}(x)$
\STATE $\emax(y), \emin(y), e_y \gets \texttt{coarsened\_quantize}(y)$
\STATE $\texttt{prod} \gets \max_j \max\bigl(\emax_j(x) + \emin_j(y),\ \emin_j(x) + \emax_j(y)\bigr)$
\STATE $\Einf^{(c)} \gets \max(e_x + e_y - \texttt{prod},\ 0)$ \COMMENT{$0$ if $x$ or $y$ is all zeros}
\IF{$53 + \Einf^{(c)} + 1 > 4000$}
\RETURN fallback
\ENDIF
\RETURN $53 + \Einf^{(c)} + 1$
\end{algorithmic}
\end{algorithm}

\subsubsection{Binned ESC}

While coarsened ESC provides a safe bound for $\esc_\infty$, scientific applications often multiply sufficiently sparse matrices to trigger this ambiguous path, as we will show later.  We now propose an alternate computational method for bounding $\esc_\infty$ using a floating-point binning algorithm. Our new algorithm is designed to be efficiently computed using fast, low-precision bfloat16 tensor cores.

The core idea is to let a lower precision floating-point dot product handle the exponent bookkeeping. Let $\Fbf$ be the set of bfloat16 values. We quantize elements $x\in F$ into nonnegative bin representatives $b(x) \in \Fbf$ based on their exponent distance from the maximum exponent in their vector. Each bin covers a fixed range of exponent space; the first holds elements within a few binades of the maximum, and each subsequent bin holds the next range below it. Unlike the coarsened ESC algorithm, we are no longer able to calculate ESC values across the full double precision exponent range, due to the much smaller exponent range of bfloat16. As a result, we use the final bin for non-zero elements that are too small to fit into the compressed exponent space. During quantization, each non-zero element is replaced with a canonical representative for the bin, typically the smallest value in the range covered by the bin. Zero inputs are mapped to zero bfloat16 representatives. Algorithm~\ref{alg:binned_quantize} describes this mapping in detail.

The binned ESC algorithm is parameterized by three meta-parameters \texttt{in\_width}, \texttt{in\_top}, and \texttt{out\_width}:
\begin{itemize}
\item \textbf{\texttt{in\_width}} is the number of binades in $F$ the are represented within a bin.  Larger values increase the maximum ESC value that can be successfully bounded but also increase the overprediction over the exact value of $\esc(x,y)$ (see equation~\eqref{e:binned_overprediction}).
\item \textbf{\texttt{in\_top}} is a special value for the number of binades in $F$ that are represented within the most significant bin.  A larger \texttt{in\_top} also influences overprediction and the maximum ESC value that can be calculated.
\item \textbf{\texttt{out\_width}} is the number of binades separating bins in $\Fbf$.  A smaller value increases the number of bins but can diminish the final algorithm's performance due to more frequent exponent update steps (smaller \texttt{k\_max} in Algorithm \ref{alg:esc_binned}).
\end{itemize}

For Ozaki-II with double-precision inputs, the configuration $\texttt{in\_width} = 4, \texttt{in\_top} = 2, \texttt{out\_width} = 8$ maximizes ESC resolution while supporting a maximum ESC of 100 or more.  To achieve good performance with $\texttt{out\_width}$ to 8, we implemented a specialized binned ESC kernel which uses bfloat16 tensor cores and periodically updates maximum exponents and resets the accumulator within the kernel.

Unlike the coarsened ESC algorithm, the overprediction in the binned ESC algorithm can be bounded by a constant. The binning provides an exact exponent span capacity after rounding into a bin, so the only error comes from combining multiple exponents into a single bin. The maximum exponent adjustment is $\texttt{in\_width} - 1$ per element or rounding up a trivial ESC value, leading to a worst-case overprediction
\begin{equation}
    \label{e:binned_overprediction}
    \esc_{\text{binned}}(x,y) - \esc(x,y) \le 2 \max(\texttt{in\_top}, \texttt{in\_width} - 1) + 1.
\end{equation}

\begin{algorithm}[tp]
\caption{\texttt{binned\_quantize}$(x)$}\label{alg:binned_quantize}
\begin{algorithmic}[1]
\REQUIRE $x \in F^k$; \texttt{in\_width}, \texttt{in\_top}, \texttt{out\_width}, \texttt{top}, \texttt{nbins}
\ENSURE $\texttt{q} \in F^k$, powers of two (or zero)
\STATE $\texttt{exp}_i \gets \floor{\log_2 \abs{x_i}}$
\STATE $\texttt{loss}_i \gets \max_j \texttt{exp}_j - \texttt{exp}_i$
\STATE $\texttt{bin}_i \gets \min\bigl(\ceil{\max(\texttt{loss}_i - \texttt{in\_top}, 0)/\texttt{in\_width}},\, \texttt{nbins}\bigr)$
\STATE $\texttt{q}_i \gets 2^{\texttt{top} - \texttt{out\_width}\cdot\texttt{bin}_i}$ if $x_i \neq 0$, else $0$
\RETURN \texttt{q}
\end{algorithmic}
\end{algorithm}

\begin{algorithm}[tp]
\caption{Binned ESC estimate for $x, y \in F^k$}\label{alg:esc_binned}
\begin{algorithmic}[1]
\REQUIRE $x, y \in F^k$; \texttt{in\_width}, \texttt{in\_top}, \texttt{out\_width}
\ENSURE $E \ge \esc_{\inf}(x, y)$
\STATE $\texttt{top} \gets \floor{(127 - \texttt{out\_width})/2}$
\STATE $\texttt{nbins} \gets \floor{(\texttt{top} + 75)/\texttt{out\_width}}$
\STATE $\texttt{k\_max} \gets 2^{\texttt{out\_width}} - 1$
\STATE $\texttt{q\_x} \gets \texttt{binned\_quantize}(x)$
\STATE $\texttt{q\_y} \gets \texttt{binned\_quantize}(y)$
\STATE $\texttt{dot} \gets 0$
\FOR{$\texttt{start} = 1, 1 + \texttt{k\_max}, 1 + 2\,\texttt{k\_max}, \dots, k$}
  \STATE $\texttt{chunk} \gets \texttt{start}, \dots, \min(\texttt{start} + \texttt{k\_max} - 1,\, k)$
  \STATE $\texttt{dot} \gets \max\bigl(\texttt{dot},\, \sum_{i \in \texttt{chunk}} \texttt{q\_x}_i \, \texttt{q\_y}_i\bigr)$
\ENDFOR
\IF{$\texttt{dot} = 0$}
  \RETURN $0$
\ENDIF
\STATE $\texttt{b} \gets \max\bigl(0,\, \ceil{(2\,\texttt{top} - \floor{\log_2 \texttt{dot}})/\texttt{out\_width}}\bigr)$
\IF{$\texttt{b} \ge \texttt{nbins} - 1$}
  \RETURN fallback
\ENDIF
\RETURN $2\,\texttt{in\_top} + \texttt{in\_width}\cdot\texttt{b}$
\end{algorithmic}
\end{algorithm}

\subsection{Computing $p$-Exponent Span Capacity}\label{sec:alg-cest}

To compute a $p$-norm based exponent span capacity for finite $p\ge 1$, we notice that $\abs{x^p} \cdot \abs{y^p} = \norm w_p^p$, where $\abs{x^p}$ is shorthand for the vector $(\abs{x_1}^p,\dotsc,\abs{x_k}^p)$. Therefore, as long we can get an approximation of $\tilde{z}^p \in \Fbf$ such that $\tilde{z} \le\abs{x^p} \cdot \abs{y^p}$, then we know our calculated $\esc_p$ is a safe over-estimation.  Similar to the binned infinity-norm exponent span capacity, we need to carefully handle the reduced range.  Both of these considerations can be handled through quantization.  In practice, $p=1$ or $p=2$ are the most relevant.

We reuse binned ESC's \texttt{exponent\_top} and \texttt{exponent\_bottom} constants, which are derived from the underlying datatype, and we align our elements to the \texttt{element\_top} and the maximum exponent for each vector.  Any non-zero element that is too small for this compressed range is stored with the minimum exponent \texttt{exponent\_bottom} and requires an extra step in our algorithm to ensure that the output is not influenced by these small elements.  If they are, we dispatch to the same sentinel as in the binned ESC algorithm. The detailed steps can be found in Algorithm~\ref{alg:esc_p}.

There are three sources of error in Algorithm~\ref{alg:esc_p}:
\begin{enumerate}
\item[(a)] rounding error in the quantization step $\tilde x_i\gets \mathrm{rd}_{\texttt{bf16}}(\abs{x_i}^p)$;
\item[(b)] rounding error from finite precision accumulation in the bfloat16 multiplication $\tilde z=\tilde x\cdot\tilde y$.
\item[(c)] limited exponent range, causing entries of $x$, $y$ that are very small relative to the maximum to become inflated.
\end{enumerate}
For (a), the use of rounded-down arithmetic ensures that $\tilde x_i$, $\tilde y_i$ are underestimates of their exact values, and hence $E_p$ is an overestimate. In theory (b) can be handled similarly, but hardware low-precision matrix multiplication accelerators typically do not allow for configurable rounding modes. However, bfloat16 matrix multiplication units typically accumulate in IEEE-754 single precision, so that we can bound the accumulator relative error by $2^{-24}k$, and multiply $\tilde z$ by $1-2^{-24}k$ to ensure that $\tilde z$ is an underestimate of its exact value. To handle (c), the final check on $\ell$ ensures that there are enough contributions from non-underflowed values to deduce a valid bound on $\esc_p$.

In the special case $p = 1$, we can also calculate an upper bound on the number of integer bits (\texttt{c\_bits}) required to represent the product $z$ exactly (Algorithm~\ref{alg:esc_1}). This is useful for Ozaki-II, where this bound can help us choose an appropriate number of moduli--- see Section~\ref{s:scaling} for more details. Note that to avoid underestimating \texttt{c\_bits}, which could lead to catastrophic accuracy loss, Algorithm~\ref{alg:esc_1} rounds upwards, which can result in slightly underestimating $\esc_1$. It is possible to avoid underestimation by carefully evaluating the effects of the rounding error and correcting for them after forming the dot product $\tilde z$.

When calculating \texttt{c\_bits} for a full matrix multiplication, care must be taken when reducing results across dot products.  Rather than element-wise calculating $\hat M$ and \texttt{c\_bits}, then maximizing across the dot products, it is necessary to first find the maximum $\hat M$ and \texttt{extra\_c\_bits} values from Algorithm~\ref{alg:esc_1} for each dot product, then calculate $\texttt{c\_bits} = 2 \hat M + \texttt{extra\_c\_bits}$.

\begin{algorithm}[tp]
\caption{\texttt{esc\_p\_quantize}$(x)$}\label{alg:esc_p_quantize}
\begin{algorithmic}[1]
\REQUIRE $x \in F^k$; $e_x$, \texttt{exponent\_top}, \texttt{exponent\_bottom}, \texttt{rounding\_function}
\IF{$x = 0$}
\RETURN $0$
\ENDIF
\STATE $\tilde x \gets \texttt{rounding\_function}\bigl(\abs{x}^p\, 2^{\texttt{exponent\_top} - p e_x}\bigr)$
\RETURN $\max(\tilde x, 2^{\texttt{exponent\_bottom}})$
\end{algorithmic}
\end{algorithm}

\begin{algorithm}[tp]
\caption{$\esc_p$ calculation for $x, y \in F^k$, $1 \le p < \infty$}\label{alg:esc_p}
\begin{algorithmic}[1]
\REQUIRE $x, y \in F^k$ finite; \texttt{exponent\_top}, \texttt{exponent\_bottom}
\ENSURE $\hat M$ with $E_p \ge \esc_p(x, y)$, or fall back
\STATE $e_x \gets \max_i \ceil{\log_2 \abs{x_i}}$
\STATE $e_y \gets \max_i \ceil{\log_2 \abs{y_i}}$
\STATE \COMMENT{special care is needed to avoid underflow and overflow}
\STATE $\tilde x_i \gets \texttt{esc\_p\_quantize}(x_i, e_x, \mathrm{rd}_{\texttt{bf16}})$
\STATE $\tilde y_i \gets \texttt{esc\_p\_quantize}(y_i, e_y, \mathrm{rd}_{\texttt{bf16}})$
\STATE $\tilde z \gets \sum_i \tilde x_i\, \tilde y_i$  \COMMENT{calculated in single precision}
\IF{$\tilde z = 0$}
\RETURN $53$
\ENDIF
\STATE $\ell \gets \floor{\log_2 \tilde z}$
\STATE \COMMENT{entries clamped at $2^{\texttt{exponent\_bottom}}$ may dominate $\tilde z$}
\IF{$\ell \le \texttt{exponent\_top} + \texttt{exponent\_bottom} + \ceil{\log_2 k}$}
\RETURN fallback
\ENDIF
\STATE $E_p \gets \max\bigl(\ceil{(2\,\texttt{exponent\_top} - \ell + 1)/p},\ 0\bigr)$ \COMMENT{$+1$: \texttt{ru}}
\RETURN $53 + E_p + 1$
\end{algorithmic}
\end{algorithm}

\begin{algorithm}[tp]
\caption{$\esc_1$ and accumulator bit count for $x, y \in F^k$}\label{alg:esc_1}
\begin{algorithmic}[1]
\REQUIRE $x, y \in F^k$ finite; $\texttt{exponent\_top}$, $\texttt{exponent\_bottom}$
\ENSURE $\hat M$ and \texttt{c\_bits}, or fallback
\STATE $e_x \gets \max_i \ceil{\log_2 \abs{x_i}}$
\STATE $e_y \gets \max_i \ceil{\log_2 \abs{y_i}}$
\STATE $\tilde x_i \gets \texttt{esc\_p\_quantize}(x_i, e_x, \mathrm{ru}_{\texttt{bf16}})$
\STATE $\tilde y_i \gets \texttt{esc\_p\_quantize}(y_i, e_y, \mathrm{ru}_{\texttt{bf16}})$
\STATE $\tilde z \gets \sum_i \tilde x_i\, \tilde y_i$
\IF{$\tilde z = 0$}
\RETURN $\hat M = 53$, $\texttt{c\_bits} = 2 \hat M$
\ENDIF
\STATE $\ell \gets \floor{\log_2 \tilde z}$
\IF{$\ell \le \texttt{exponent\_top} + \texttt{exponent\_bottom} + \ceil{\log_2 k}$}
\RETURN fallback
\ENDIF
\STATE $E_1 \gets \max(2\,\texttt{exponent\_top} - \ell,\ 0)$ \COMMENT{may add 1 for safety to compensate for rounding inputs up}
\STATE $\hat M \gets 53 + E_1 + 1$
\STATE $\texttt{extra\_c\_bits} \gets \max\bigl(\ceil{\log_2 \tilde z} - 2\,\texttt{top} + 1,\ 0\bigr)$ \COMMENT{$+1$: sum rounding}
\RETURN $\hat M$, $\texttt{c\_bits} = 2 \hat M + \texttt{extra\_c\_bits}$
\end{algorithmic}
\end{algorithm}

\section{ESC in Ozaki-II}
\label{s:scaling}

In this section, we briefly note how ESC-derived mantissa bit counts can be integrated into the Ozaki-II algorithm. For more detailed background on the algorithm, see~\cite{ozaki2026scheme2,caday2026twoM}. In Ozaki et al's original description~\cite{ozaki2026scheme2}, accuracy is selected via scaling factors, and the authors present several methods (``fast'' and ``accurate'' scaling) that aim to maximize input integer precision subject to a fixed accumulator precision, parameterized by modulus count.

Our viewpoint is the opposite: instead of maximizing accuracy subject to a given performance target (modulus count), we wish to maximize performance subject to the accuracy requirement embodied in the mantissa bit count $\hat M$. An additional benefit of viewing scaling this way is that as the underlying datatypes change~\cite{uchino2026fp8,HayashiFP4} we can maintain comparable, even bitwise equivalent, accuracy.

To achieve this, we first determine Ozaki-II scaling factors with a ``fixed scaling'' method. Letting $a_i$ denote the $i^{\text{th}}$ row of $A$, the scaling factor for that row is given by $\sigma_i={2}\hat\eps(\norm {a_i}_\sinfty)$. Similarly the $j^{\text{th}}$ column $b_j$ of $B$ will be scaled by $\tau_j={2}\hat\eps(\norm {b_j}_\sinfty)$. Note that $\hat\eps$ captures the mantissa bit requirement $\hat M$. This matches the fixed-point discretization scheme of Section~\ref{s:accuracy-definitions}.

After scaling and discretization, Ozaki-II multiplies its integer inputs in modular arithmetic, modulo a set of small coprime integers $m_1,\dotsc,m_N$, then applies the Chinese Remainder Theorem and lifts the result from the modular domain to an integer in the range $(-M_N/2, M_N/2)$, where $M_N = \prod_{i=1}^N m_i$. To avoid catastrophic wraparound when lifting, it is therefore necessary to choose $N$ (and therefore $M_N$) large enough that the integer product's elements lie within $(-M_N/2, M_N/2)$. That is, we require an upper bound on the elements of the product matrix $C=AB=(c_{ij})$; roughly speaking%
    \footnote{This inequality is not exact due to the fixed-point discretization.}%
, we need $\abs{c_{ij}}\lesssim \sigma_i\tau_jM_N/2$. Subject to this requirement, we wish to minimize $N$ to minimize computation time, which means bounding $C$ from above as tightly as possible.

In the worst case, we have the simple upper bound $\abs{c_{ij}}\leq \sigma_i\tau_j 2^{2\hat M}k$. A tighter bound on $C$ can be found by precalculating the 2-norms of each row of $A$ and column of $B$ and applying the Cauchy-Schwarz inequality, similar to the fast scaling method. Alternatively, if we have already calculated $\esc_1(A,B)$, the algorithm in Section~\ref{sec:alg-cest} provides this upper bound directly via its \texttt{c\_bits} output.

\section{Accuracy Studies}\label{sec:accuracy}

Let us begin by providing a name to an accuracy definition that was first used in~\eqref{e:fp-worst-jr} and described as Jeannerod and Rump\textquotesingle s ``worst-case error bound'' (the entry $\eps(1)\cdot k\,\|w\|_1$ in the row ``Floating point, worst order'' of Table~\ref{t:bounds}, and the bound that Section~\ref{s:matching-bounds} asks the fixed-point worst-case error not to exceed). \textbf{Criterion 1} (the classical bound). $|\hat z - z| \le k\,\eps(1)\,\|w\|_1$, where $\|w\|_1 = \sum_l |x_l y_l|$. We use it for two results: the fixed-point dot product $\hat z$ and, with $\mathrm{fl}(\hat z)$ in place of $\hat z$, the FP64 number that is returned (``after fl'' in Table~\ref{tab:c1-nearest}). This is the classical worst-case bound for a floating-point dot product in any summation order, Higham\textquotesingle s inner-product bound~\cite{higham2002accuracy}, $\gamma_k\,\|w\|_1$ with $\gamma_k = ku/(1 - ku)$ and unit roundoff $u = \eps(1)$, with $k \eps(1)$ in place of $\gamma_k$. The two agree to first order in $\eps(1)$. It is the requirement of Fasi, Mary, and Mikaitis~\cite{fasi2026sharper} that the emulated product satisfy this same bound (their Eq.\ (3.3) is a mantissa bit count that guarantees it). With $k = n$, the second form is also the threshold of the published grade-A tests of Demmel et al.~\cite{demmel2026grade} ($f_A(n) < n$ with $\epsilon = 2^{-53} = \eps(1)$, where $f_A(n)$ denotes the largest ratio $|\hat C_{ij} - C_{ij}| / (\epsilon\,(|A|\,|B|)_{ij})$ over the entries of the computed $\hat C$). Grade A as defined admits any polynomial $f_A(n)$. Criterion~1 measures the fixed-point error against the worst case of floating point. It does not cover the average-case comparison of Section~\ref{s:matching-bounds}, in which the fixed-point variance is to be no larger than the floating-point variance.

A \emph{rule} returns a mantissa bit count, which we also call its \emph{width}. The rules compared below are those of Section~\ref{sec:algorithms}; except for binned, a rule\textquotesingle s width is 53 plus its exponent term plus a constant $C$. The Criterion~1 ratios of Table~\ref{tab:c1-nearest} come from 463 constructed and random instances, $k$ from 2 to 2049, every rule at its prescribed width under round to nearest, exact arithmetic. The two rows of Table~\ref{tab:c1-nearest} for the rules built on the 1-norm and on the 2-norm of $w$ (Section~\ref{sec:alg-cest}) are from a second run, of 12 November, which adds three constructed instances built against those rules\textquotesingle{} base width of 56 bits and draws the random families anew. All measurements use a developer build of cuBLAS, planned for an upcoming release, with test code for the Grading the BLAS from Demmel et al. ~\cite{demmel2026grade} (test code from ~\cite{gradingblas2026code}). Table~\ref{tab:c1-nearest}, the summary of Criterion~1 over these instances, is placed in Section~\ref{sec:adversaries}, beside the inputs that attain its worst ratios.

\subsection{``Grading the BLAS''}\label{sec:grading}

\begin{figure}[htbp]\centering\includegraphics[width=0.88\textwidth]{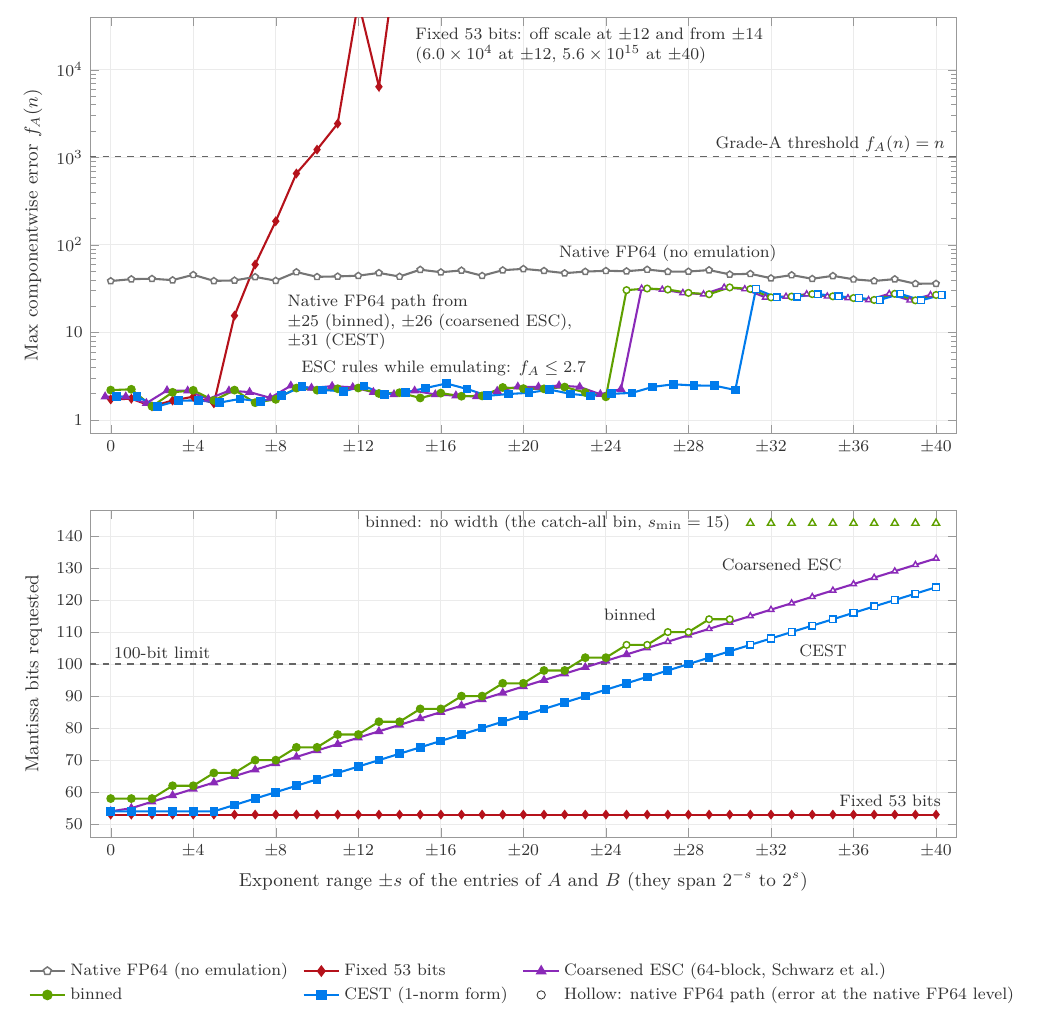}
\caption{cuBLAS on ``Grading the BLAS''~\cite{demmel2026grade} Test 2c, $m = n = k = 1024$: on an NVIDIA B200, at every exponent range from $\pm 0$ to $\pm 40$, the maximum componentwise error $f_A(n)$ of five modes, with the grade-A threshold $f_A(n) = n$, and the mantissa bits each rule requests, with the library\textquotesingle s nominal 100-bit limit and binned\textquotesingle s no-width cells. By these widths the hand-off came a few bits above that limit. The widths are those of the rules, evaluated on matrices of the test\textquotesingle s construction, not measured: these runs logged none. Every width the library logged in separate runs equals the evaluated width (Section~\ref{sec:grading}). Hollow markers: the GEMM ran in FP64, the native run throughout and a rule\textquotesingle s run where it took the FP64 path (from $\pm 25$ for binned, $\pm 26$ for the coarsened ESC, and $\pm 31$ for CEST), which its error level shows (at least 23 there, at most 2.7 while the fixed-point kernels ran). Every error is the harness\textquotesingle s exact value.}\label{fig:test2c-n1024}\end{figure}

\begin{figure}[htbp]\centering\includegraphics[width=0.95\textwidth]{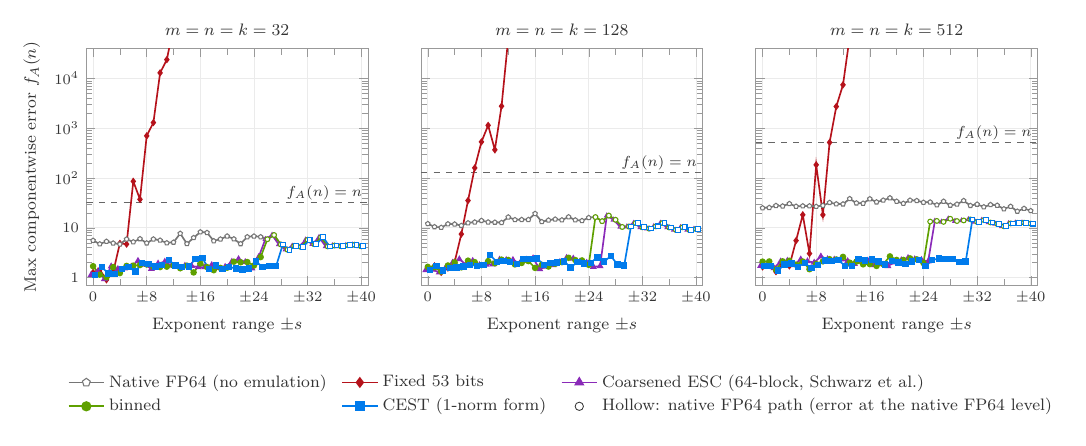}
\caption{cuBLAS on ``Grading the BLAS''~\cite{demmel2026grade} Test 2c, maximum componentwise error against the exponent range, by matrix size: the error panel of Figure~\ref{fig:test2c-n1024} for three more of the six sizes, $n = 32$, 128, and 512, each against its own grade-A line $f_A(n) = n$. At these three sizes the three rules stay below 2.9 while they emulate (the largest value, 2.83, is CEST at $n = 128$, $\pm 9$) and at or above 3.6 on the FP64 path; binned hands off at $\pm 25$ ($\pm 26$ at $n = 32$), the coarsened ESC at $\pm 26$ ($\pm 27$ at $n = 128$), and CEST at $\pm 28$, $\pm 30$, and $\pm 31$ as $n$ grows. For $n \le 128$ the FP64 path returns the native run\textquotesingle s product, so from the hand-off the native curve lies under the rules\textquotesingle{} hollow markers.}\label{fig:test2c-all-n}\end{figure}

\begin{figure}[htbp]\centering\includegraphics[width=0.95\textwidth]{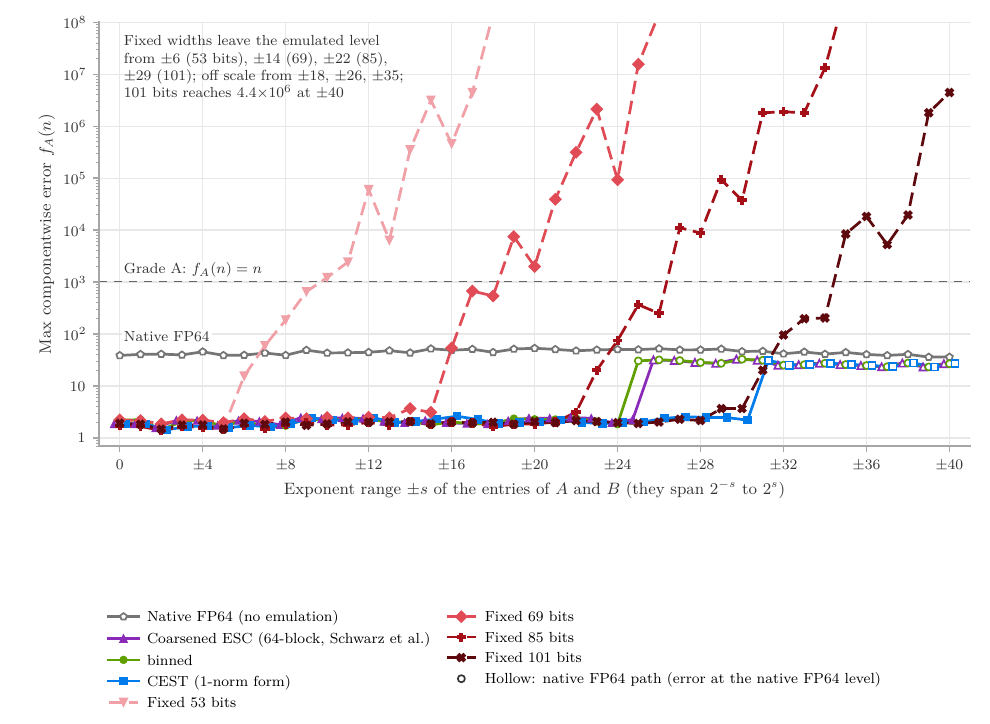}
\caption{cuBLAS on ``Grading the BLAS''~\cite{demmel2026grade} Test 2c, $m = n = k = 1024$, fixed widths against the width rules: on an NVIDIA B200, the maximum componentwise error of the runs at fixed mantissa widths of 53, 69, 85, and 101 bits (dashed) against native FP64 and the three width rules of Figure~\ref{fig:test2c-n1024}. Each fixed width leaves the emulated level where the exponent range outgrows it, first at $\pm 6$, $\pm 14$, $\pm 22$, and $\pm 29$, and first loses grade A at $\pm 10$, $\pm 19$, $\pm 27$, and $\pm 35$ (the climb is not monotone from range to range); 53 bits is off the scale from $\pm 18$ ($5.6 \times 10^{15}$ at $\pm 40$), 69 from $\pm 26$, 85 from $\pm 35$, and 101 bits reaches $4.4 \times 10^{6}$ at $\pm 40$.}\label{fig:test2c-fixed}\end{figure}

\begin{figure}[htbp]\centering\includegraphics[width=0.88\textwidth]{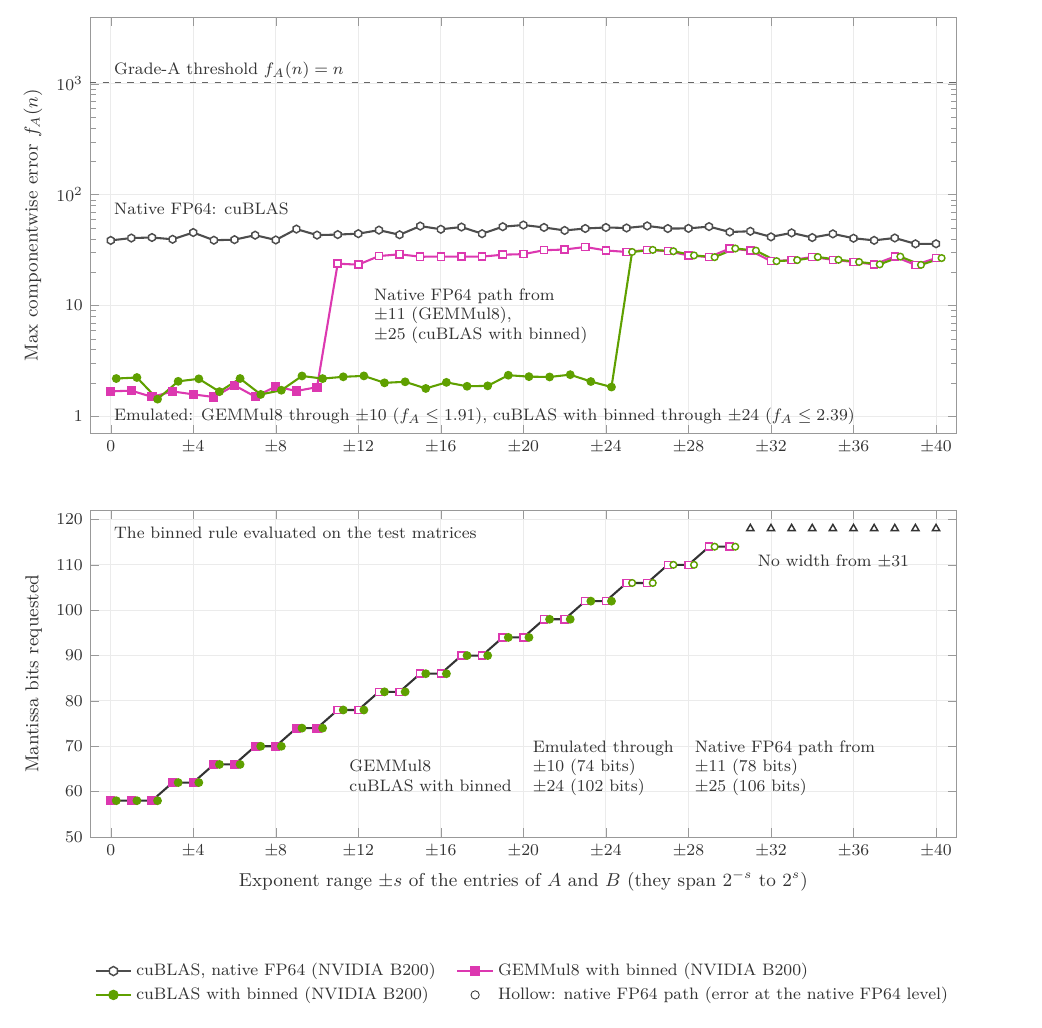}
\caption{GEMMul8 on an NVIDIA B200, ``Grading the BLAS''~\cite{demmel2026grade} Test 2c, $m = n = k = 1024$: GEMMul8 with the binned rule, with two of the cuBLAS runs of Figure~\ref{fig:test2c-n1024} (cuBLAS with the same rule and its native FP64 GEMM); the maximum componentwise error $f_A(n)$ at every exponent range from $\pm 0$ to $\pm 40$, with the grade-A threshold $f_A(n) = n$, and the mantissa bits binned requests, each width carrying the marker of GEMMul8 and that of cuBLAS with binned (triangles: the ranges, from $\pm 31$, at which the rule gives no width). GEMMul8 logs no width, and these cuBLAS runs logged none: the widths are those of the rule, evaluated on matrices of the test\textquotesingle s construction, the same widths as binned\textquotesingle s in Figure~\ref{fig:test2c-n1024}. Hollow markers: the GEMM ran in FP64, the native run throughout, GEMMul8 from $\pm 11$, and cuBLAS with binned from $\pm 25$, which the error level of each of the two shows (for GEMMul8 at least 23 from $\pm 11$ and at most 1.91 through $\pm 10$; for cuBLAS with binned at least 23 from $\pm 25$ and at most 2.39 through $\pm 24$). From $\pm 25$ GEMMul8 and cuBLAS with binned return the same product; the markers of cuBLAS with binned are set a little to the right of their ranges so that both runs show.}\label{fig:gemmul8-n1024}\end{figure}

The Grading the BLAS tests grade BLAS implementations through tests designed to be ``progressively less gameable.''  Where the goal is to achieve an ``A'' grade, where the published grade-A threshold is Criterion 1 with $k = n$ (the start of Section~\ref{sec:accuracy}), so every rule that meets Criterion 1 after the final rounding to FP64 meets this bar.  Most relevant to the study of exponent span capacity is Test 2, which iteratively increases the dynamic range under permutation and sign changes and was designed specifically to be adversarial for fixed-point methods.  Test 2 consists of three increasingly more challenging subtests 2a, 2b, and 2c of which our measurements in Figures \ref{fig:test2c-n1024}, \ref{fig:test2c-all-n}, \ref{fig:test2c-fixed}, and \ref{fig:gemmul8-n1024} use test 2c.

Figures \ref{fig:test2c-n1024} and \ref{fig:test2c-all-n} demonstrate that a fixed 53 mantissa bit Ozaki-II implementation is not sufficient to achieve a grade A. The high dynamic range of the inputs along with the narrow dynamic range of the fixed-point representation leads to many elements rounding away many or even all of their significant bits.  This leads to much higher max componentwise error than native FP64.  All of the exponent span capacity methods dynamically adapt the number of mantissa bits and consistently handle the increased dynamic range from the tests.  The plots also show that the automatic dynamic precision framework dispatches to native FP64 once the dynamic range is sufficiently large (in this case roughly 100 mantissa bits, indicated by a hollow symbol in the plots) and obtain a grade A.  When the Ozaki-II scheme is used, the max componentwise error is often more than an order of magnitude smaller than the native FP64.

One may wonder if we can get away with just increasing the number of fixed mantissa bits past 53 and start obtaining a grade A.  While possible, to do this robustly you would need to maintain a potentially impractical number (in excess of 2000) mantissa bits.  Finally, Figure \ref{fig:test2c-fixed} shows several fixed mantissa bit counts running on test 2c and that the dynamic range of test 2c eventually becomes too large for them to be graded an A.

CEST denotes the rule built on the 1-norm of $w$ (Section~\ref{sec:alg-cest}), with exponent term $E_1 = e_x + e_y - \lfloor \log_2 \|w\|_1 \rfloor$ ($e_x$ and $e_y$ as in Section~\ref{sec:alg-esc}), the integer form of $\esc_1(x, y)$ (the exponent of each norm in place of its logarithm). Its width is $53 + E_1 + 3$ with a floor of 54 bits, the form of the widths the library logged. Algorithm~\ref{alg:esc_1} takes $e_x$ and $e_y$ rounded up, as $\lceil \log_2 \|x\|_\infty \rceil$ and $\lceil \log_2 \|y\|_\infty \rceil$: where no entry is clamped at $2^{\texttt{exponent\_bottom}}$, its $E_1$ is never more than one less than this $E_1$, and it is this $E_1$ plus 1 or 2 when neither maximum is a power of two and the $\max(\cdot, 0)$ of its line 13 does not bind, so that its width, 53 plus its own $E_1$ plus 1, is then the width above or one bit less.\par \textbf{The runs.} Nine runs of Test 2c of the ``Grading the BLAS'' harness~\cite{demmel2026grade} (BLAS, the Basic Linear Algebra Subprograms) against cuBLAS on an NVIDIA B200 are behind Figures~\ref{fig:test2c-n1024} to~\ref{fig:test2c-fixed}: native FP64 without emulation, the three width rules the library implements (binned, CEST, and the coarsened ESC), and fixed widths of 53, 55, 69, 85, and 101 bits (the figures draw all but the 55-bit run), each at the six sizes $m = n = k = 32$, 64, \dots, 1024 and at every exponent range from $\pm 0$ to $\pm 40$, the entries of $A$ and $B$ spanning $2^{-s}$ to $2^{s}$ at range $\pm s$. These runs logged no width. Under a rule the library either emulates the GEMM (the fixed-point kernels run) or takes the FP64 path (its fallback to FP64 arithmetic). Which of the two happened is read from the result: for $n \le 128$ the FP64 path returns the native run\textquotesingle s product bit for bit, and for $n \ge 256$, where its product differs from the native run\textquotesingle s, its error $f_A(n)$ (the start of Section~\ref{sec:accuracy}) is above 7 and the emulated error is below 3. The widths shown with the runs (Figure~\ref{fig:test2c-n1024}) are those of the three rules, in the forms given in the caption of Figure~\ref{fig:synth-bits}, evaluated on matrices of the test\textquotesingle s construction; in separate runs with the library\textquotesingle s log on, that evaluation agreed with the log in all 306 cases (283 widths, and 23 cases in which binned logged none).

\textbf{The width rules.} The three rules earn grade A at every size and range. While they emulate, $f_A(n)$ stays below 3.1 (the largest value, 3.01, is CEST at $n = 64$, $\pm 16$; at $n = 1024$ it is at most 2.7). Native FP64 earns grade A throughout as well, with $f_A$ from 3.6 to 53 across the sizes, growing with $n$ as a sum of $n$ roundings should (its largest value is 8.2 at $n = 32$ and 53 at $n = 1024$). Fixed 53 bits does not: it is better than native at no range beyond $\pm 5$ to $\pm 9$, depending on the size, first loses grade A between $\pm 6$ and $\pm 10$, and from $\pm 31$ its error is of the order of $|A|\,|B|$ itself, with $f_A$ between $1.9 \times 10^{15}$ and $5.7 \times 10^{15}$ (Figures~\ref{fig:test2c-n1024} and \ref{fig:test2c-all-n}).

\textbf{The hand-off to FP64.} Each rule emulates up to a range and runs every wider range on the FP64 path. The binned rule emulates through $\pm 24$ ($\pm 25$ at $n = 32$), the coarsened ESC through $\pm 25$ or $\pm 26$, and CEST through $\pm 27$ to $\pm 30$, depending on the size. The library\textquotesingle s nominal cap on the emulated width is 100 bits (the 100-bit limit of Figure~\ref{fig:test2c-n1024}) and can be extended upward; in these runs, by the evaluated widths, the hand-off came a few bits past it. On the FP64 path the three rules return the same product at every size.

\textbf{Fixed mantissa widths.} The runs at 53, 69, 85, and 101 bits, with a run at 55 bits that the figures do not draw, measure what a fixed request delivers as the range grows, the inverse of the question the rules answer. Each fixed width keeps the emulated error level, $f_A$ below 3.1, up to a range that grows with the width, and beyond it the error climbs, though not monotonically from range to range. The error first passes 3.1 at $\pm 4$ to $\pm 6$ for 53 bits, depending on the size, at $\pm 6$ or $\pm 7$ for 55, at $\pm 12$ or $\pm 14$ for 69, at $\pm 20$ to $\pm 22$ for 85, and at $\pm 29$ or $\pm 30$ for 101. Grade A, $f_A < n$, is first lost one to six ranges later, the three largest sizes holding it longer than the three smallest: at $\pm 6$ to $\pm 10$ for 53 bits, $\pm 7$ to $\pm 12$ for 55, $\pm 15$ to $\pm 19$ for 69, $\pm 24$ to $\pm 27$ for 85, and $\pm 32$ to $\pm 35$ for 101, each further 16 bits of width moving the loss out by about eight ranges (Figure~\ref{fig:test2c-fixed} shows $n = 1024$). At 101 bits the errors stay at the emulated level through $\pm 28$ at every size, where native FP64 has $f_A$ from 39 to 53 at $n = 1024$, so the fixed-point kernels do run at a 101-bit request.

\textbf{Fixed widths and grade A.} Demmel et al.\textquotesingle s~\cite{demmel2026grade} remark that with enough slices or moduli nothing prevents an A is borne out and bounded. A fixed width earns grade A on Test 2c only as far as the exponent range lets it (the thresholds above and Figure~\ref{fig:test2c-fixed}), while the width rules earn it at every size and range by choosing the width per call and taking the FP64 path past the cap.

\textbf{GEMMul8 on the same matrices.} Figure~\ref{fig:gemmul8-n1024} shows GEMMul8 with the binned rule on the same Test 2c matrices at $n = 1024$, run on an NVIDIA B200, beside two of the runs of Figure~\ref{fig:test2c-n1024}, cuBLAS with the same rule and the native FP64 GEMM of cuBLAS. GEMMul8 emulates through $\pm 10$, with $f_A$ at most 1.91, and runs in FP64 from $\pm 11$, where binned requests 78 bits (74 at $\pm 10$); cuBLAS with the same rule emulates through $\pm 24$, with $f_A$ at most 2.39, and runs in FP64 from $\pm 25$, where binned requests 106 bits (102 at $\pm 24$).

\subsection{Random data}\label{sec:random}

The \emph{exponent rules} are the exponent-sum ESC (uncoarsened, $c = 1$ in~\eqref{eq:esc-coarsened}), with exponent term $\Einf = \Einf^{(1)}$, and the coarsened ESC; the \emph{norm forms} are CEST and CEST$'$, which is CEST with the 2-norm of $w$ in place of the 1-norm, so that its exponent term is $E_2 = e_x + e_y - \lfloor \log_2 \|w\|_2 \rfloor$. At the proved widths (see Section~\ref{sec:proof-c1}) the random families (fine mantissa bits, flat, lognormal, random signs) stay inside the Criterion 1 bound, below a ratio of $0.4$ (the worst, $0.39$, from the fine-mantissa-bit family at 54 bits, the width of the exponent rules there), where the constructed adversary reaches $0.999$; we therefore lead with constructed inputs and report random results as the typical case.

\begin{figure}[htbp]\centering\includegraphics[width=0.95\textwidth]{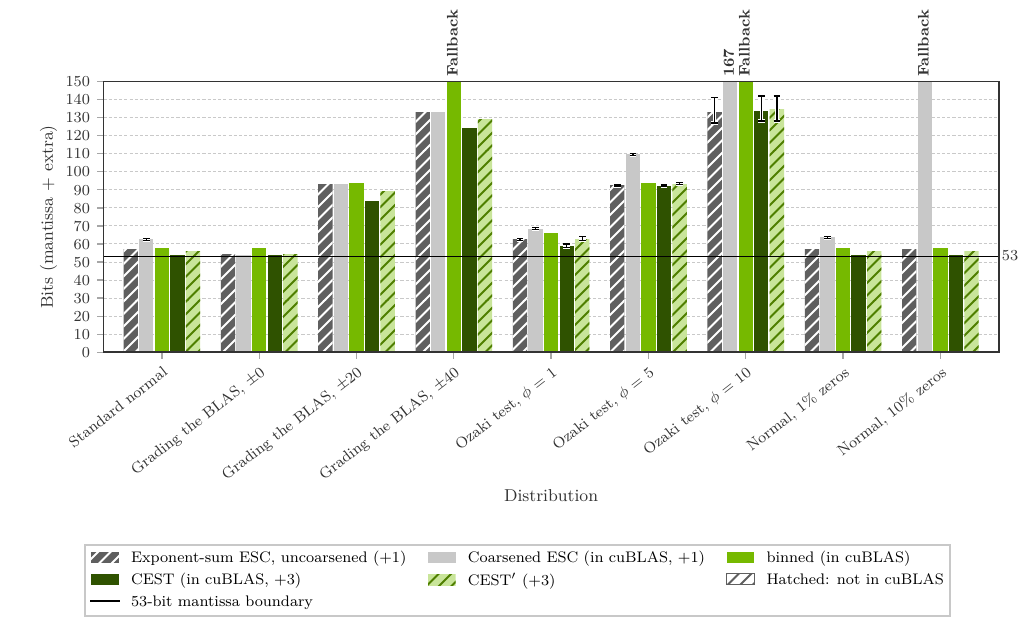}
\caption{Synthetic distributions ($m = n = k = 1024$), all values in bits: mantissa bits requested by the five rules for a GEMM of that size on synthetic matrices, calculated from the rules of this paper (Section~\ref{sec:algorithms}) on the matrices themselves, not with cuBLAS: no bar is a width measured in the library. One width per GEMM, the largest over its $n^2$ dot products. Bars: the mean over five draws of $A$ and $B$; whiskers: the smallest and largest draw. Grading the BLAS: Test 2c~\cite{demmel2026grade} at exponent range $\pm s$, one construction each. Ozaki test: entries $(\mathrm{rand} - 0.5) \exp(\phi\,\mathrm{randn})$~\cite{ozaki2026scheme2}. Normal, 1\% and 10\% zeros: standard normal entries, each set to zero with that probability. The three rules that are in cuBLAS take the form of the widths it logged (Section~\ref{sec:grading}): the coarsened ESC at $53 + \Einf^{(c)} + 1$, binned at $58 + 4 s_{\min}$ ($s_{\min}$ is the bin \texttt{b} of Algorithm~\ref{alg:esc_binned}), and CEST at $53 + E_1 + 3$ with a floor of 54. CEST$'$, which is not in cuBLAS, takes the same form with $E_2$. The exponent-sum ESC is $53 + \Einf + 1$. These five forms are the rules of Table~\ref{tab:c1-nearest}. Fallback: the rule gives no width. The bar marked 167 is off the scale (162 to 175 bits over the draws). The chart shows requests only: in the runs of Section~\ref{sec:grading} cuBLAS ran in FP64 every request that, by the evaluated widths, was more than a few bits above its nominal 100-bit cap.}\label{fig:synth-bits}\end{figure}

\textbf{Bits requested on synthetic matrices.} Figure~\ref{fig:synth-bits} shows the mantissa bits the five rules request for a GEMM with $m = n = k = 1024$ on nine kinds of synthetic matrices. The bars are calculated from the rules of Section~\ref{sec:algorithms} on the matrices themselves, in the forms the caption gives; none is a width measured with cuBLAS. On standard normal entries the rules ask 54 to 63 bits: CEST 54, CEST$'$ 56, the exponent-sum ESC 57, binned 58, and the coarsened ESC 62 or 63, since it bounds the largest product of a block of 64 from below, by the largest exponent of one vector in the block plus the smallest of the other. One percent of zeros moves only the coarsened ESC, to 63 or 64 bits. With ten percent the coarsened ESC has no width and the other four are unchanged; at this size the coarsened ESC loses its width between one and two percent of zeros. On Test 2c the two exponent rules agree (54, 93, and 133 bits at $\pm 0$, $\pm 20$, and $\pm 40$); binned asks one bit more at $\pm 20$ and has no width at $\pm 40$, and at both CEST asks nine bits fewer and CEST$'$ four. On matrices of the form that Ozaki, Uchino, and Imamura use as tests~\cite{ozaki2026scheme2} (they take $\phi$ from 0.5 to 4) the rules ask 58 to 69 bits at $\phi = 1$; at $\phi = 5$ four of them ask 92 to 94 bits and the coarsened ESC 109 or 110; at $\phi = 10$ binned has no width, the coarsened ESC asks 162 to 175 bits, and the other three 127 to 142.

\subsection{Adversarial details}\label{sec:adversaries}

Each Criterion 1 constant of the exponent rules comes with an input that attains it; the norm forms are at the $+3$ of the library, two bits above what Criterion~1 needs before the final rounding (Section~\ref{sec:proof-c1}).

At a width of $\hat M$ bits the entries of $x$ are rounded to the grid of step $2^{e_x + 1 - \hat M} = 2\hat\eps(\|x\|_\infty)$, and those of $y$ likewise with $e_y$. The FP64 unit in the last place (ulp) of a nonzero real $a$ is $2\eps(a)$, and $y_l$ is the \emph{partner} of $x_l$.\par \textbf{The midpoint adversary}: an input on which rounding the converted operands to nearest moves every fixed-point term up by nearly half a grid step, so that the error is the most the width allows. Here $x_1 = y_1 = 1$, a pair in the top binade of both vectors, so $\Einf = 0$ and every rule sits at its base width. The other $k - 1$ entries of $x$ lie one FP64 ulp above a midpoint $(m_l + \tfrac12)\,g$ of the $\hat M$-bit grid, $g = 2^{1-\hat M}$ (since $e_x = 0$), between $2^{20-\hat M}$ and $2^{21-\hat M}$ (33 to 34 binades below the anchor $x_1 = y_1 = 1$ at $\hat M = 54$). Their partners in $y$ are $2 - 4 \cdot 2^{-52}$, just below 2, so that each term\textquotesingle s error is multiplied by nearly 2. Nearest rounding at $\hat M$ bits moves every small entry up by almost half a step, so $\hat z$ lands $(k-1)\,2^{53-\hat M}$ ulps of $z$ above $z$, $(k-1)/2$ ulps at $\hat M = 54$, against an allowance $k\,\eps(1)\,\|w\|_1$ of about $k/2$ ulps. It is generated by construction, not by search: for the target count $\hat M$, draw an integer $m_l$ for each small entry uniformly from $[2^{19}, 2^{20}]$ (in steps of $2^{\hat M-54}$ when $\hat M > 54$, so that each midpoint lies $2^{-\hat M}$ above a point of the 54-bit grid), set $x_l$ to the FP64 number one ulp above $(m_l + \tfrac12)\,g$ and $y_l = 2 - 4 \cdot 2^{-52}$, and verify the claim by exact rational arithmetic, computing $z$ and the fixed point at every width. The random $m_l$ only avoid a special pattern, and any values in the range give the same numbers.

\textbf{Criterion 1 holds at every rule\textquotesingle s own width}, with a worst ratio of $0.999$ over the instances, that is $(k-1)/k$, to the allowance at $k = 1024$, attained by the midpoint adversary above on the exponent rules at their width of 54 bits (Table~\ref{tab:c1-nearest}). For binned the worst ratio is $0.06$, one sixteenth of that, as its width is four bits more. For the norm forms, at $+3$ with a floor of 54, it is $0.25$, from the same adversary built against their 56 bits. For the exponent rules $(k-1)/k$ is also the worst case over all inputs, and for the norm forms the ratio is below $1/4$ on every input (Section~\ref{sec:proof-c1}). After the fixed-point result is rounded to FP64 the worst ratio over the instances is unchanged for the exponent rules, $0.25$ for the norm forms, and $0.16$ for binned, the one cell of Table~\ref{tab:c1-nearest} set by a random input, from the fine-mantissa-bit family at $k = 7$, where half an ulp of final rounding weighs against an allowance of only a few ulps. After the rounding Criterion~1 holds on every input for the exponent rules and for the norm forms at $+3$; at $+1$ the norm forms would exceed the allowance on inputs built for the purpose (ratios of 1.25 and 1.50 at $k = 2$), which is why they are not used at $+1$ (Section~\ref{sec:proof-c1}).

\textbf{Mixed signs} inherit every single-sign case: no mixed-sign cell exceeds its single-sign value.

\begin{table}[htbp]\centering\small\setlength{\tabcolsep}{5pt}\renewcommand{\arraystretch}{1.15}
\caption{Criterion 1 under round to nearest: every rule at its own width, $53 + C$ plus the row\textquotesingle s own exponent term ($\Einf$, $E_1$, or $E_2$), with $C$ its constant. Green, safe: the criterion holds at the rule\textquotesingle s own width on every instance, and the number in parentheses is the worst ratio $|\hat z - z| / (k\,\eps(1)\,\|w\|_1)$ over the instances, the error against the Criterion 1 allowance, which is $k/2$ to $k$ ulps of $\|w\|_1$; 1 would be the bound itself. The norm forms are at $+3$ with a floor of 54 bits, the form of the widths cuBLAS logged for CEST (Section~\ref{sec:grading}); at $+1$ the rounding to FP64 could take their result past the allowance (Section~\ref{sec:proof-c1}). For binned, $s_{\min}$ is the bin \texttt{b} of Algorithm~\ref{alg:esc_binned}. ``After fl'' is the same ratio after the fixed-point result is rounded to FP64. Mixed-sign inputs inherit every cell. Every cell by exact rational arithmetic.}\label{tab:c1-nearest}
\begin{tabular}{@{}l c c@{}}\toprule
Rule (constant under round to nearest) & Criterion 1 & Criterion 1 after fl \\ \midrule
Exponent-sum ESC, uncoarsened ($+1$) & \cellcolor[HTML]{E4F1D0}Safe (0.999) & \cellcolor[HTML]{E4F1D0}Safe (0.999) \\
Coarsened ESC, 64-block ($+1$) & \cellcolor[HTML]{E4F1D0}Safe (0.999) & \cellcolor[HTML]{E4F1D0}Safe (0.999) \\
binned ($58 + 4 s_{\min}$) & \cellcolor[HTML]{E4F1D0}Safe (0.06) & \cellcolor[HTML]{E4F1D0}Safe (0.16) \\
1-norm form, CEST ($+3$, floor 54) & \cellcolor[HTML]{E4F1D0}Safe (0.25) & \cellcolor[HTML]{E4F1D0}Safe (0.25) \\
2-norm form, CEST$^{\prime}$ ($+3$, floor 54) & \cellcolor[HTML]{E4F1D0}Safe (0.25) & \cellcolor[HTML]{E4F1D0}Safe (0.25) \\
\bottomrule\end{tabular}
\end{table}

The \emph{exponent span} of the data is the number of binades between the smallest and the largest nonzero entry exponents.\par \textbf{The wasted maxima} (any rule, any span), the range family parametrized by $s$: entries confined to $[2^{-s}, 2^{s+1})$, a span of $2s$ binades, the maximum of $x$ paired with the minimum of $y$ and vice versa, the remaining entries of $x$ at the floor of the span, one FP64 ulp above a midpoint of the $\hat M$-bit grid so that nearest rounding moves each by almost half a step, and the remaining entries of $y$ at the top of the span. This is within one bit of the worst Criterion 1 input for a given exponent span: it needs $2s + 43$ bits at $k = 1024$ (no input of the span needs more than $2s + 44$, which an input with both entries of each remaining index at the floor of the span needs for $s \le 8$), against which the exponent-sum ESC and the coarsened ESC provide $2s + 54$, binned $2s + 58$ ($2s + 56$ at odd $s$; no width from $s = 30$, as the rule falls back from bin 15, line 15 of Algorithm~\ref{alg:esc_binned}), CEST (the 1-norm form) $2s + 46$, and CEST$'$ (the 2-norm form) $2s + 51$ (each raised to 54 where it is less). Under the library\textquotesingle s nominal 100-bit cap (Section~\ref{sec:grading}) the exponent rules request more than the cap from $s = 24$ and binned from $s = 22$, while at $s = 24$ CEST and CEST$'$ still request less (94 and 99 bits against a true need of 91); on Test 2c, by the evaluated widths, the hand-off came a few bits past the cap. Test 2 of Demmel et al.~\cite{demmel2026grade} (Section~\ref{sec:grading}) has the same structure, but its $y$ entries are not all at the top, so it needs less than this family at the same $s$, by 6 to 12 bits for $s$ up to 40 and by 10 to 12 at $s = 100$, 200, \dots, 500, the last being the range of the published Tests 2b and 2c ($2^{\pm 500}$); the counts are for one draw of the test\textquotesingle s vector and move by a few bits with the draw.

{%
\textbf{The shortest instances, in decimal.} Each pattern above is already present in a vector of three or four entries, and the entries are ordinary FP64 numbers, so the instances can be printed (Tables~\ref{tab:midpoint-instance} and~\ref{tab:wasted-instance}) and typed into any implementation. Every decimal in the two tables is the shortest string that reads back to the same FP64 number under round to nearest (Python\textquotesingle s \texttt{repr}), so the vectors reproduce bit for bit, and the form columns give the construction.

\emph{The midpoint adversary at $k = 4$} (Table~\ref{tab:midpoint-instance}; built against $\hat M = 54$). The three small entries lie one FP64 ulp above a midpoint of the 54-bit grid, $g = 2^{-53}$, and their partners $2 - 2^{-50}$ are just below 2, so the fixed point lands $1.5$ ulps above $z$ against an allowance of 2 ulps, a ratio of $0.75$ $= (k-1)/k$, and Criterion 1 holds at 54 bits. At binned\textquotesingle s 58 bits the midpoints of the 54-bit grid are grid points and the error is negligible, about $3.5 \times 10^{-10}$ ulps of $z$ (the entries lie one FP64 ulp above those points). Appending a negligible negative product, $x_5 = 2^{-90}$ and $y_5 = -1$, changes neither the widths nor the fixed point, so the ratio is $0.6 = (k-2)/k$ at $k = 5$, the single-sign inheritance in its smallest form.

\emph{The wasted maxima at $k = 3$, $s = 4$} (Table~\ref{tab:wasted-instance}). The maximum of $x$ meets the minimum of $y$ and vice versa, and the third pair is the floor of the span against its top: the largest and the smallest product are one binade apart, yet $\Einf = 2s = 8$ and the rules ask 62 (the exponent-sum ESC and its block coarsening), 66 (binned), 62 (CEST), and 63 (CEST$'$) bits. Every entry is represented exactly from 61 bits on.

\begin{table}[htbp]\centering\small
\caption{The midpoint adversary under round to nearest, built against $\hat M = 54$, at $k = 4$: the entries by their form and as the shortest decimals that read back to the same FP64 numbers.}\label{tab:midpoint-instance}
{\footnotesize\begin{tabular}{@{}r l l l l@{}}\toprule
$l$ & $x_l$ (form) & $y_l$ (form) & $x_l$ & $y_l$ \\ \midrule
1 & $1$ & $1$ & 1 & 1 \\
2 & $(m + \tfrac12)\,2^{-53} + 2^{-86}$, $m = 754715$ & $2 - 2^{-50}$ & 8.379025251414875e-11 & 1.9999999999999991 \\
3 & $(m + \tfrac12)\,2^{-53} + 2^{-86}$, $m = 874262$ & $2 - 2^{-50}$ & 9.70626357066351e-11 & 1.9999999999999991 \\
4 & $(m + \tfrac12)\,2^{-53} + 2^{-86}$, $m = 956121$ & $2 - 2^{-50}$ & 1.0615081036391417e-10 & 1.9999999999999991 \\
\bottomrule\end{tabular}}
\end{table}

\begin{table}[htbp]\centering\small
\caption{The wasted maxima at $s = 4$, $k = 3$: the entries by their form and as the shortest decimals that read back to the same FP64 numbers.}\label{tab:wasted-instance}
{\footnotesize\begin{tabular}{@{}r l l l l@{}}\toprule
$l$ & $x_l$ (form) & $y_l$ (form) & $x_l$ & $y_l$ \\ \midrule
1 & $2^{4}$ & $2^{-4}$ & 16 & 0.0625 \\
2 & $2^{-4}$ & $2^{4}$ & 0.0625 & 16 \\
3 & $2^{-4} + 2^{-56}$ & $2^{5} - 2^{-48}$ & 0.06250000000000001 & 31.999999999999996 \\
\bottomrule\end{tabular}}
\end{table}

\textbf{How the requirement scales with $k$.} Criterion 1 does not scale. The proved (see Section~\ref{sec:proof-c1}) constants hold for every $k$, and for the exponent rules the worst case approaches the bound from below as $(k-1)/k$ ($0.75$, $0.857$, $0.984$, $0.999$ at $k = 4$, 7, 64, 1024, respectively), because the adversary\textquotesingle s $k - 1$ half-steps and the allowance $k\,\eps(1)\,\|w\|_1$ grow together; for the norm forms at $+3$ the ratio stays below $1/4$. Two $k$ dependences run the other way: the flat-data saving of the norm forms, about $\log_2 k$ and $\tfrac12 \log_2 k$ bits (Section~\ref{s:accuracy-matching}), which the floor of 54 bits caps, and the range stress, where the need falls with $k$ while the maxima-tied widths do not. The wasted maxima need about $2s + 53 - \log_2 k$ (the $2s + 43$ above at $k = 1024$) while the exponent rules\textquotesingle{} $2s + 54$ does not move. On Test 2 at $\pm 0$, $\pm 10$, and $\pm 29$ the need falls by one to two bits per doubling of $n$ (46 to 37, 63 to 55, and 99 to 91, for $n = 32$ to 1024) while every maxima-tied width moves by at most a few bits with $n$, so the rules\textquotesingle{} margin there is about $\log_2 n$ bits plus the structure\textquotesingle s own slack. Finally, $k$ enters the cost: the integer accumulator of an Ozaki-II product at $\hat M$ bits needs $2\hat M + \lceil \log_2 k \rceil + 1$ bits, one more per doubling of $k$.
}

\section{Proofs}\label{s:proofs}

In this section, we provide proofs of the error bounds in Section~\ref{sec:accuracy-theory}.

\subsection{Fundamental Uncertainty Bounds}
\label{s:uncertainty-proof}

\begin{proof}[Proof of Proposition~\ref{p:input-rounding-worst}]
Applying the inequalities $\abs{\overline x_i-x_i}\leq \eps(x_i)$, $\abs{\overline y_i-y_i}\leq \eps(y_i)$,
\begin{align}
\abs{\overline x\cdot\overline y-x\cdot y}
& \leq \sum_{i=1}^k \abs{\overline x_i\overline y_i - x_iy_i}\\
& \leq \sum_{i=1}^k \abs*{(\overline x_i - x_i)y_i + x_i(\overline y_i-y_i) + (\overline x_i - x_i)(\overline y_i - y_i)}\\
& \leq \sum_{i=1}^k \left[\eps(x_i)\abs{y_i} + \abs{x_i}\eps(y_i) + \eps(x_i)\eps(y_i)\right]\\
& \leq \sum_{i=1}^k 2\eps(1)\abs{x_i}\abs{y_i} + O(\eps(1)^2) \\
& = 2\eps(1)\norm w_1 + O(\eps(1)^2). \qedhere
\end{align}
\end{proof}

\begin{proof}[Proof of Proposition~\ref{p:input-rounding-average}]
By direct calculation, using the independence of the $X_i$, $Y_i$:
\begin{align}
\Var(X\cdot Y) = \sum_{i=1}^k \Var(X_iY_i) &= \sum\left[x_i^2 \Var(Y_i)+y_i^2\Var(X_i) + \Var(X_i)\Var(Y_i)\right]\\
&= \sum\left[\frac{1}{3}\left(x_i^2\eps(y_i)^2+y_i^2\eps(x_i)^2\right)+\frac{1}{9}\eps(x_i)^2\eps(y_i)^2\right].
\intertext{Since $\frac12\eps(1)\cdot\abs a<\eps(a)\leq \eps(1)\cdot\abs a$,}
\Var(X\cdot Y)&\geq\sum\left[\frac{1}{6}x_i^2y_i^2\left(\eps(1)^2+\frac{\eps(1)^4}{24}\right)\right]\\
&\geq\frac{1}{6}\eps(1)^2\cdot \norm w_2^2.
\end{align}
Taking square roots, the standard deviation is bounded below by
\begin{equation}
\label{e:uncertainty-lower}
\sigma(X\cdot Y) \geq \frac{1}{\sqrt{6}}\cdot \eps(1)\cdot\norm w_2.
\end{equation}
The upper bound can be derived similarly:
\begin{equation}
\label{e:uncertainty-upper}
\sigma(X\cdot Y) \leq \frac{2}{\sqrt{6}}\sqrt{1 + \frac{\eps(1)^2}{6}}\cdot \eps(1)\cdot\norm w_2.
\end{equation}
\end{proof}

\subsection{Fixed-Point Error Bounds}
\label{s:fixed-point-error-proof}

\begin{proof}[Proof of Proposition~\ref{p:fixed-worst}]
The fixed-point discretization errors can be expressed in terms of the round-off error $\hat\eps(a)\coloneq 2^{\floor{\log_2 a} - \hat M}$, applied to the largest elements $\norm x_\sinfty$, $\norm y_\sinfty$:
\begin{align}
\abs{x_i-\hat x_i} \leq \hat\eps(\norm x_\sinfty),
\qquad
\abs{y_i-\hat y_i} \leq \hat\eps(\norm y_\sinfty).
\end{align}
Plugging these errors into the dot product:
\begin{align}
\abs{\hat z - z} &\leq \sum_{i=1}^k \abs{\hat x_i\hat y_i-x_iy_i} \\
&\leq \sum\left[\hat\eps(\norm x_\sinfty)\cdot\abs{y_i} + \hat\eps(\norm y_\sinfty)\cdot\abs{x_i} + \hat\eps(\norm x_\sinfty)\hat\eps(\norm y_\sinfty)\right]\\
&=\hat\eps(\norm x_\sinfty)\cdot\norm y_1 + \hat\eps(\norm y_\sinfty)\cdot\norm x_1 + k\hat\eps(\norm x_\sinfty)\hat\eps(\norm y_\sinfty).
\end{align}
Both inequalities above, while quite pessimistic in the average case, can be made tight with an appropriate choice of input data $x$, $y$, if $E < 0$. Weakening the bound slightly further, we can pull out $\hat\eps(1)=2^{-\hat M}$:
\begin{align}
\label{e:fxp-error}
\abs{\hat z - z} &\leq \hat\eps(\norm x_\sinfty)\cdot\norm y_1 + \hat\eps(\norm y_\sinfty)\cdot\norm x_1 + O(\hat\eps(1)^2)\notag\\
    & \leq \hat\eps(1)\cdot\left[\norm x_\sinfty\norm y_1 + \norm y_\sinfty\norm x_1\right] + O(\hat\eps(1)^2).
\end{align}
We can fine-tune this bound slightly if $E > 0$, i.e., the fixed-point representation has more mantissa bits as the floating-point representation. In this case, the largest elements in $x$, $y$ can be exactly represented in fixed point, and do not contribute any error to $\hat z$. That is, $\hat x_i = x_i$ if $\abs{x_i} \geq 2^{-E}\norm x_\sinfty$, and similarly for $y$.

Let $I_x\coloneq\set{i\in\{1,\dotsc, k\}}{\abs{x_i}\geq 2^{-E}\norm x_\sinfty}$ be the set of error-free indices in $x$, and define $I_y$ similarly. For each $i$, consider the error term $r_i$ associated with discretizing $x_i$, $y_i$:
\[
r_i = (\hat x_i-x_i)\cdot{y_i} + (\hat y_i - y_i)\cdot{x_i} + O(\hat\eps(1)^2).
\]
If $i\in I_x$, then the first term of $r_i$ vanishes; similarly, if $i\in I_y$ the second term vanishes. In either case, $\abs{r_i}$ can be bounded to first order by $\hat\eps(1)\norm x_\sinfty\norm y_\sinfty$. Finally, if $i\notin I_x\cup I_y$, then $\abs{x_i}\leq\frac12\norm x_\sinfty$, $\abs{y_i}\leq\frac12\norm y_\sinfty$, and again to first order $\abs{r_i}$ is bounded by $\hat\eps(1)\norm x_\sinfty\norm y_\sinfty$. Summing over $i$, we conclude
\begin{align}
\label{e:fxp-error-finer-infty}
\abs{\hat z - z}\leq \hat\eps(1)\cdot k\norm x_\sinfty \norm y_\sinfty + O(\hat\eps(1)^2),
\end{align}
completing the proof.
\end{proof}

\begin{proof}[Proof of Proposition~\ref{p:fixed-avg}]
The proof is very similar to that of Proposition~\ref{p:input-rounding-average}. By direct calculation,
\begin{align*}
\Var(\hat x\cdot \hat y) = \sum_{i=1}^k \Var(\hat x_i\hat y_i) &= \sum\left[x_i^2 \Var(\hat y_i)+y_i^2\Var(\hat x_i) + \Var(\hat x_i)\Var(\hat y_i)\right]\\
&= \sum\left[\frac{1}{3}\left(x_i^2\hat \eps(\norm y_\sinfty)^2+y_i^2\hat \eps(\norm x_\sinfty)^2\right)+\frac{1}{9}\hat \eps(\norm x_\sinfty)^2\hat \eps(\norm y_\sinfty)^2\right]\\
&= \frac13\left[\norm x_2^2 \hat \eps(\norm y_\sinfty)^2+\norm y_2^2 \hat \eps(\norm x_\sinfty)^2\right] + \frac k9\hat \eps(\norm x_\sinfty)^2\hat \eps(\norm y_\sinfty)^2  \\
&\leq \frac{\hat \eps(1)^2}{3}\left[\norm x_2^2\norm y_\sinfty^2 + \norm y_2^2\norm x_\sinfty^2\right] + O(\hat \eps(1)^4).
\end{align*}
Taking square roots, the standard deviation is bounded above by
\begin{align}
\label{e:fxp-avg-ub}
\sigma(\hat x\cdot \hat y) \leq \frac{\hat \eps(1)}{\sqrt 3} \sqrt{\norm x_2^2\norm y_\sinfty^2 + \norm y_2^2\norm x_\sinfty^2} + O(\hat \eps(1)^2).
\end{align}
A lower bound can be extracted similarly, leading to the same right-hand side as~\eqref{e:fxp-avg-ub}, but multiplied by $\frac12$. This establishes~\eqref{e:fxp-avg}, and~\eqref{e:fxp-avg-inf} is an immediate consequence.

Note that, just as in the worst-case analysis, we may fine-tune this bound when $E>0$ by noting that the largest terms of $x$ and $y$ have no discretization error.
\end{proof}

\subsection{Floating-Point Error Bounds}
\label{s:fp-error-proof}

\begin{proof}[Proof of Proposition~\ref{p:fp-worst}]
Without loss of generality, assume $z\geq 0$. (If $z<0$, we may replace $x$ by $-x$ and negate at the end without changing the result.) Choose a permutation $\pi$ such that $w_{\pi(1)}\geq w_{\pi(2)}\geq \cdots \geq w_{\pi(k)}$.

For brevity, let us reorder the input vectors $x$, $y$ by $\pi$ so that $w_1\geq w_2\geq\dotsc\geq w_k$. Let $N_+=\abs{\set{i}{w_i>0}}$ count the number of positive $w_i$. Note that by our assumptions on $x$, $y$, the $w_i$ are all nonzero.

Let $s_i=\sum_{j=1}^i w_j$ be the vector of partial sums. The $s_i$ form a sequence that is first increasing ($i\leq N_+$), then potentially decreasing ($i > N_+$). Let us count how many $s_i$ exceed $\norm w_1/4$.

Note that $s_{N_+}\geq \norm w_1 / 2$, as
\[
\norm w_1 = \sum_{i\leq N_+} w_i + \sum_{i>N_+} -w_i = 2s_{N_+} - z \leq 2s_{N_+}.
\]
By a similar argument, $s_{\ceil{N_+/2}} \geq \norm w_1/4$ and $s_{k-\ceil{(k - N_+)/2}}\geq \norm w_1/4$. Thus $s_i \geq \norm w_1/4$ for all $i\in[\ceil{N_+/2}, k-\ceil{(k-N_+)/2}]$, which contains at least $k/2$ indices.
We conclude that
\[
\norm s_2^2 \geq \frac{k}{32} \norm w_1^2,
\]
finishing the proof.
\end{proof}

\begin{proof}[Proof of Proposition~\ref{p:fp-avg}]
Averaging the variance~\eqref{e:pfpdp-variance} over all permutations,
\begin{align}
\label{e:avg-perm-var-1}
\frac{1}{k!}\sum_\pi\Var(d_\pi(x,y)) = \frac{c\eps(1)^2}{3k!}\sum_\pi\sum_{i=1}^k\bigg(\sum_{j=1}^i w_{\pi(j)}\bigg)^2,
\end{align}
for some $c\in(\frac14, 1]$.

We will simplify equation~\eqref{e:avg-perm-var-1} in two steps. First, let us count how many times a particular instance of the innermost sum occurs. Each instance of the innermost sum is equal to $\sum_{i\in I} w_i$ for some non-empty index set $I\subseteq[k]$. Conversely, given a subset $I$, we encounter the sum $\sum_{i\in I} w_i$ in~\eqref{e:avg-perm-var-1} exactly $\abs I!\cdot(k-\abs I)!$ times, so
\begin{align}
\frac{1}{k!}\sum_\pi\Var(d_\pi(x,y)) &= \frac{c\eps(1)^2}{3k!}\sum_{I\subseteq[k]}\bigg(\sum_{i\in I}w_i\bigg)^2\cdot\abs I!\cdot(k-\abs I)!\\
&= \frac{c\eps(1)^2}{3}\sum_{I\subseteq[k]}\bigg(\sum_{i\in I}w_i\bigg)^2\cdot{k\choose\abs I}^{-1}.
\label{e:avg-perm-var-2}
\end{align}
After expanding the square, we will get a linear combination of binomials $w_iw_j$. Separate these into two classes: cross-terms ($w_iw_j$, $i\neq j$) and squares ($w_i^2$). Even without knowing the coefficients of each binomial, we know that by symmetry all the cross-terms have the same coefficient, say $Q_1$, and all the squares have identical coefficients, say $Q_2$:
\[
\frac{1}{k!}\sum_\pi\Var(d_\pi(x,y)) = c\eps(1)^2\Bigg[\sum_{1\leq i<j\leq k} Q_1w_iw_j + \sum_{i=1}^k Q_2w_i^2\Bigg].
\]
Changing variables to $Q_1=2N_1$, $Q_2=N_1 + N_2$, this simplifies to
\[
\frac{1}{k!}\sum_\pi\Var(d_\pi(x,y)) = c\eps(1)^2\left[N_1(\textstyle\sum w_i)^2 + N_2\norm w_2^2\right] = c\eps(1)^2\left[N_1z^2 + N_2\norm w_2^2\right].
\]
All that remains is to calculate the coefficients $N_1$, $N_2$. Letting $\ell=\abs I$ and counting the number of times a cross-term $w_iw_j$ appears in~\eqref{e:avg-perm-var-2},
\begin{align*}
2N_1 = \frac23\sum_{\ell = 2}^k {k\choose\ell}^{-1}{k-2\choose\ell-2}
	= \frac23\sum_{\ell = 2}^k \frac{\ell(\ell - 1)}{k(k-1)}
	&= \frac2{3k(k-1)}\sum_{\ell = 2}^k \left[\ell(\ell - 1)\right]\\
	&= \frac2{3k(k-1)}\frac{k(k+1)(k-1)}{3}\\
	&= \frac29(k+1).
\end{align*}
The square terms are simpler:
\begin{align*}
N_1+N_2=\frac13\sum_{\ell=1}^k{k\choose\ell}^{-1}{k-1\choose\ell-1}=\frac13\sum_{\ell=1}^k\frac{\ell}{k}=\frac16(k+1),
\end{align*}
and we conclude $N_1=(k+1)/9$, $N_2=(k+1)/18$. Therefore the average variance over all permutations is
\begin{align}
\frac{1}{k!}\sum_\pi\Var(d_\pi(x,y)) = \frac{(k+1)c\eps(1)^2}{18}\cdot\left[2z^2+\norm w_2^2\right],
\end{align}
completing the proof.
\end{proof}

\subsection[The Criterion 1 constants]{The Criterion 1 constants}\label{sec:proof-c1}

For Criterion~1, the first line of~\eqref{e:fxp-error} (in the proof of Proposition~\ref{p:fixed-worst}), with $\|y\|_1 < k\,2^{e_y + 1}$, gives $+2$ for every rule whose exponent term satisfies $2^{e_x + e_y - E} \le \|w\|_1$, which all four do (the $+3$ of Fasi, Mary, and Mikaitis~\cite{fasi2026sharper} is for a full-step conversion error). The $+1$ of this paper holds by a sharper argument, index by index. In this paragraph let $E$ be the exponent term of a rule (elsewhere $E = \hat M - M$) with $2^{e_x + e_y - E} \le \|w\|_1$, let its width satisfy $\hat M \ge 54 + E$ and $\hat M \ge 53$, and put $B = 2^{e_x + e_y + 1 - \hat M}$, so that $B \le \eps(1)\,\|w\|_1$. An FP64 entry $x_i$ is a multiple of $2^{e(x_i) - 52}$, so on the grid of step $2^{e_x + 1 - \hat M}$ it converts exactly unless it lies at least $\hat M - 52$ binades below its vector\textquotesingle s maximum, and its conversion error is at most $2^{e_x - \hat M}$. An index with one inexact entry therefore errs by less than $2^{e_x - \hat M}\,2^{e_y + 1} = B$, and an index with two inexact entries, both that far below their maxima, by less than $B$ as well, the step that needs $\hat M \ge 53$. The $k$ indices together err by less than $kB \le k\,\eps(1)\,\|w\|_1$, which is Criterion~1 for the fixed-point result at $+1$. For the exponent rules ($E \ge \Einf$) the index that attains the largest exponent sum converts exactly, so the error is below $(k-1)\,\eps(1)\,\|w\|_\infty$; the midpoint adversary reaches $(k-1)/k$ to within $10^{-6}$ (Section~\ref{sec:adversaries}), and the constant is tight for $k \ge 3$: built against one bit fewer, the same adversary has the ratio $2(k-1)/k$. At $+1$ the norm forms would be covered as well, since $E_1$ satisfies the inequality through the floor of $\log_2 \|w\|_1$ and $E_2$ through $\|w\|_2 \le \|w\|_1$ (with widths raised to 53 where the formula gives less); but for them every index can err, and the ratio comes within $10^{-12}$ of 1 for CEST (at $k = 2$, 4, and 1024) and within $10^{-6}$ of 1 for CEST$'$ (at $k = 2$, 4, and 64). The final rounding of $\hat z$ to FP64 adds up to $\eps(1)\,|\hat z|$. The exponent rules keep Criterion~1 after it, the exact index leaving room for that half ulp. The norm forms at $+1$ do not: on $x = (2 - 2^{-50},\, a)$, $y = (a,\, 2 - 5 \cdot 2^{-52})$ with $a = 2^{-8}(1 + 3 \cdot 2^{-52})$, CEST at $+1$ asks 60 bits, and the ratio is just below 1 before that rounding and 1.25 after it (1.50 on an input on which CEST$'$ at $+1$ asks 60 bits as well). The norm forms are therefore used at $+3$ with a floor of 54, the form of the widths the library logged for CEST (Section~\ref{sec:grading}), here extended to CEST$'$: then $B \le \eps(1)\,\|w\|_1 / 4$, so the ratio is below $1/4$ before the final rounding and below $1/4 + 1/k$ after it for $k \ge 2$ (at $k = 1$ both entries convert exactly), and Table~\ref{tab:c1-nearest} shows them at that width. The tightened bound of Proposition~\ref{p:fixed-worst} does not supply the $+1$: with $\|x\|_\infty \|y\|_\infty < 2^{e_x + e_y + 2}$ it is below $k\,2^{e_x + e_y + 2 - \hat M}$ to first order, the limit of the first line of~\eqref{e:fxp-error}, so it gives $+2$ as well. Kept in the norms, it is the condition~\eqref{e:worst-match} of Section~\ref{s:matching-bounds}, $\hat M - 53 \ge \max(1, \esc_1(x, y))$; since $E_1 - 1 < \esc_1(x, y) < E_1 + 2$, the smallest width that meets it is $53 + E_1$, $53 + E_1 + 1$, or $53 + E_1 + 2$ where $E_1 \ge 1$, and, as that bound is stated, it gives Criterion~1 to first order in $\hat\eps(1)$.

\section{Related Work}\label{sec:related}

Two papers derive the bit count for FP64 emulation on integer tensor cores from the exponent spread. Schwarz et al.~\cite{schwarz2026guaranteed} give the exponent-sum ESC with a $+1$ margin for the mantissa-product carry, coarsened over blocks along $k$. Their paper is generic in the block length, while the library\textquotesingle s coarsened ESC uses 64. Fasi, Mary, and Mikaitis~\cite{fasi2026sharper} supply the rigorous derivation, whose sharpest bound, their Eq.\ (3.3), is the 1-norm form with $+1$ on exact logarithms, and whose max-form bound carries $+3$. Their separate widths for $x$ and $y$ are a possible further saving not evaluated here. Of the roughly twenty papers citing the two papers, none replaces ``ESC plus a constant''. Abdelfattah, Dongarra, Fasi, Mikaitis, and Tisseur~\cite{abdelfattah2025analysis} analyze the error from the input side with a per-row factor and note that it may fail on badly scaled rows. The Ozaki-II line~\cite{ozaki2026scheme2} fixes $\hat M = 53$ (Kouya~\cite{kouya2026cpus} adds 16 guard bits) and adapts the modulus count instead, with at least 14 INT8 moduli for FP64 (Uchino~\cite{uchino2026fp8}). Mazumder, Calotoiu, and Hoefler~\cite{mazumder2026fp64} keep truncation and observe that the vendor\textquotesingle s path sets its width per call, which a fixed count cannot. Matsuoka~\cite{matsuoka2026ozaki25} proposes an ESC-adaptive modulus count and notes that truncation does not inherit the round-to-nearest bound, so it must be validated separately. The density-fitting paper~\cite{huang2026density} picks its slice count from the self-consistent-field (SCF) energy change with no formal bound, and the other application papers (device simulation~\cite{maeder2026semiconductor}, eigenvector refinement~\cite{terao2026eigenvector}, mixed-precision singular value decomposition (SVD)~\cite{schwarz2026svd}) adopt the method without a rule of their own. Caday~\cite{caday2026twoM} observes that the Fasi, Mary, and Mikaitis constant on the max form (ESC $+3$, his $q^*$) consistently overpredicts, which agrees with our $+2/+1$ for that form, that nearest rounding is worth a bit, and that ESC $+2$ suffices for his 2M complex GEMM. Demmel et al.~\cite{demmel2026grade} make the bit count an explicit open problem. This paper adds proved (in Section~\ref{sec:proof-c1}) and attained constants under round to nearest, coarsening safety, and the zero and range conventions.

\section{Conclusion}\label{sec:conclusions}

We have analyzed the worst-case and average-case error of fixed-point and floating-point dot products and, by matching the fixed-point bounds to the floating-point bounds, obtained the $p$-exponent span capacities. Of these, $\esc_1$ suffices to match the worst-case error of floating point and $\esc_2$ its average-case error, so the original ESC is conservative for both. The capacities can be approximated with a low-precision (bfloat16) GEMM, and the binned ESC algorithm approximates the original ESC more robustly, and often more tightly, than the coarsened ESC does. In Ozaki-II, the mantissa bit count fixes the scaling factors, and the $\esc_1$ algorithm also returns a bound on the product from which the number of moduli can be chosen (Section~\ref{s:scaling}).

For the exponent rules and the norm forms, this paper proves that choosing the mantissa bit count automatically is numerically safe in the sense of Criterion~1, the classical worst-case bound for a floating-point dot product, and for these rules and binned it demonstrates that safety in software. With the constants of Section~\ref{sec:proof-c1}, $+1$ for the exponent rules and $+3$ with a floor of 54 bits for the norm forms, these rules keep Criterion~1 on every input, before and after the final rounding to FP64, and the constant of the exponent rules is tight for $k \ge 3$. In exact arithmetic, each of the five rules of Table~\ref{tab:c1-nearest} kept the bound at its own width on every instance, constructed or random, the midpoint adversary reaching $(k-1)/k$ of the allowance on the exponent rules. In a developer build of cuBLAS on an NVIDIA B200, the three rules that the library implements earned grade~A on Test~2c of ``Grading the BLAS''~\cite{demmel2026grade} at all six sizes and at every exponent range from $\pm 0$ to $\pm 40$. A fixed count of 53 bits first lost the grade between $\pm 6$ and $\pm 10$, depending on the size. Each rule emulated up to a range, beyond which the library ran the GEMM in FP64, and the grade was kept in both cases.

One possible further saving remains to be evaluated: the separate widths of Fasi, Mary, and Mikaitis~\cite{fasi2026sharper} for $x$ and $y$. The cuBLAS measurements used a developer build, which is planned for an upcoming release.

\end{document}